\documentstyle [12pt,preprint,eqsecnum,aps,amsfonts] {revtex}
\tighten
\input epsf
\newcommand{\beq}{\begin{equation}}
\newcommand{\eeq}{\end{equation}}
\newcommand{\beqs}{\begin{eqnarray}}
\newcommand{\eeqs}{\end{eqnarray}}

\draft
\widetext

\begin{document}

\baselineskip 6.0mm

\preprint{YITP-SB-99-58, BNL-HET-99/32}

\title{Ground State Entropy of the Potts Antiferromagnet with 
Next-Nearest-Neighbor Spin-Spin Couplings on Strips of the Square Lattice}

\vspace{4mm}

\author{
Shu-Chiuan Chang$^{(a)}$\thanks{email: shu-chiuan.chang@sunysb.edu},
Robert Shrock$^{(a,b)}$\thanks{email: robert.shrock@sunysb.edu. On sabbatical
leave at BNL; permanent address: (a)}}

\address{(a) \ C. N. Yang Institute for Theoretical Physics \\
State University of New York \\
Stony Brook, N. Y. 11794-3840, USA}

\address{(b) \ Physics Department \\
Brookhaven National Laboratory \\
Upton, NY  11793-5000}

\maketitle

\vspace{10mm}

\begin{abstract}

We present exact calculations of the zero-temperature partition function
(chromatic polynomial) and $W(q)$, the exponent of the ground-state entropy,
for the $q$-state Potts antiferromagnet with next-nearest-neighbor spin-spin
couplings on square lattice strips, of width $L_y=3$ and $L_y=4$ vertices and
arbitrarily great length $L_x$ vertices, with both free and periodic boundary
conditions.  The resultant values of $W$ for a range of physical $q$ values are
compared with each other and with the values for the full 2D lattice.  These
results give insight into the effect of such non-nearest neighbor couplings on
the ground state entropy.  We show that the $q=2$ (Ising) and $q=4$ Potts
antiferromagnets have zero-temperature critical points on the $L_x \to \infty$
limits of the strips that we study.  With the generalization of $q$ from
${\mathbb Z}_+$ to ${\mathbb C}$, we determine the analytic structure of $W(q)$
in the $q$ plane for the various cases.

\end{abstract}


\vspace{16mm}

\pagestyle{empty}
\newpage

\pagestyle{plain}
\pagenumbering{arabic}
\renewcommand{\thefootnote}{\arabic{footnote}}
\setcounter{footnote}{0}

\section{Introduction}

The $q$-state Potts antiferromagnet (AF) with the usual nearest-neighbor
spin-spin couplings \cite{potts,wurev} exhibits nonzero ground state entropy,
$S_0 > 0$ (without frustration) for sufficiently large $q$ on a given lattice
$\Lambda$.  This is equivalent to a ground state degeneracy per site $W > 1$,
since $S_0 = k_B \ln W$.  Such nonzero ground state entropy is important as an
exception to the third law of thermodynamics \cite{al,chowwu}.  A physical
example of nonzero ground state entropy is ice \cite{lp}-\cite{liebwu}. 
In this $q$-state Potts
antiferromagnet at $T=0$, the value of each spin must be different than the
values of all of the other spins to which it is coupled.  
There is a close connection with graph theory here,
since the zero-temperature partition function of the above-mentioned $q$-state
Potts antiferromagnet on a lattice $\Lambda$ or, more generally, a graph $G$,
satisfies 
\beq 
Z(G,q,T=0)_{PAF}= P(G,q)
\label{zp}
\eeq
where $P(G,q)$ is the chromatic polynomial expressing the number of ways of 
coloring the vertices of the graph $G$ with $q$ colors such that no two 
adjacent vertices have the same color (for reviews, see 
\cite{rrev}-\cite{bbook}).   The minimum number of colors for which this
coloring is possible, i.e. the minimum integer value of $q$ for which $P(G,q)$
is nonzero, is denoted the chromatic number of $G$, $\chi(G)$. 

 From eq. (\ref{zp}), it follows 
that\footnote{\footnotesize{At certain special
points $q_s$ (typically $q_s=0,1,.., \chi(G)$), one has the noncommutativity of
limits $\lim_{q \to q_s} \lim_{n \to \infty} P(G,q)^{1/n} \ne \lim_{n \to
\infty} \lim_{q \to q_s}P(G,q)^{1/n}$, and hence it is necessary to specify the
order of the limits in the definition of $W(\{G\},q_s)$ \cite{w}.  We use the
first order of limits here; this has the advantage of removing certain isolated
discontinuities in $W$.}}
\beq
W(\{G\},q) = \lim_{n \to \infty} P(G,q)^{1/n}
\label{w}
\eeq 
where $n=v(G)$ is the number of vertices of $G$ and $\{G\} = \lim_{n \to
\infty}G$.  

Since $P(G,q)$ is a polynomial, one can generalize $q$ from
${\mathbb Z}_+$ to ${\mathbb C}$.  The zeros of $P(G,q)$ in the complex $q$
plane are called chromatic zeros; a subset of these may form an accumulation
set in the $n \to \infty$ limit, denoted ${\cal B}$, which is the continuous
locus of points where $W(\{G\},q)$ is nonanalytic. The maximal region in the
complex $q$ plane to which one can analytically continue the function
$W(\{G\},q)$ from physical values where there is nonzero ground state entropy
is denoted $R_1$.  The maximal value of $q$ where ${\cal B}$ intersects the
(positive) real axis is labelled $q_c(\{G\})$.  This value is important since 
$W(\{G\},q)$ is a real analytic solution for real $q$ down to $q_c(\{G\})$.
For regions other than $R_1$, one can only determine the magnitude
$|W(\{G\},q)|$ unambiguously \cite{w}.  In addition to
\cite{rrev}-\cite{w}, some previous works on chromatic polynomials 
include \cite{bds}-\cite{t}. 

In previous works we have carried out comparative studies of $W$ for different
lattices and have explored the effects of different lattice properties such as
coordination numbers \cite{w},\cite{wn}-\cite{wn},\cite{strip}-\cite{w2d},
\cite{pg,wcy}, \cite{pm}-\cite{tw},\cite{t}.  
In general it was found that as one
increased the lattice coordination number, the ground state entropy of the
$q$-state Potts antiferromagnet (if nonzero for the given value of $q$),
decreased. This can be understood as a consequence of the fact that as one
increases the lattice coordination number, one is increasing the constraints on
the coloring of a given vertex subject to the constraint that other vertices of
the lattice adjacent to this one (i.e. connected with a bond or edge of the
lattice) have different colors.  Another way in which to explore this effect is
to consider non-nearest-neighbor spin-spin couplings.  Again, in general, these
increase the constraints on the values that any given spin can take on, and
hence decrease the ground state entropy.  We wish to make this more
quantitative and shall do so here using exact solutions for the chromatic
polynomials and rigorous bounds.  A natural starting point for studies
of such non-nearest-neighbor spin-spin couplings is to consider the model on a
given lattice and add next-nearest-neighbor (nnn) spin-spin couplings.
Equivalently, we can redefine the lattice itself by considering it as a graph
$G$ with vertices at the usual lattice sites but with bonds (= edges in graph
theory nomenclature) consisting not only of the usual bonds joining these
lattice sites but also bonds joining next-nearest-neighbor lattice sites.  We
then consider the nearest-neighbor Potts antiferromagnet on this redefined
lattice.

Perhaps the simplest case that one could consider is the $q$-state Potts
antiferromagnet in one dimension; the lattice here is just the a line $T_n$ or
circle $C_n$ for the case of free and periodic boundary conditions,
respectively (denoted $FBC_x$ and $PBC_x$).  One has $P(T_n,q)=q(q-1)^{n-1}$ so
$W=q-1$ and $R_1$ is the entire $q$ plane.  For the circle,
$P(C_n,q)=(q-1)^n+(q-1)(-1)^n$.  In this case, if $|q-1| > 1$ then $W=q-1$,
while if $|q-1| \le 1$, then $|W|=1$, so that $q_c=2$.  Hence for either
$FBC_x$ or $PBC_x$ there is nonzero ground state entropy $S_0=k_B\ln(q-1)$ for
$q > 2$.  The addition of next-nearest neighbor bonds converts $T_n$ or $C_n$
to a open or cyclic strip of triangles, respectively, with each pair sharing an
edge.  We denote these strips $tri(L_y=2),N_t,BC_x$ where $BC_x=FBC_x$ or 
$PBC_x$.  In the cyclic case, the degree $\Delta$ (number of neighboring 
vertices) of each
vertex is changed from 2 to 4, and this is also true of the internal vertices
in the open case.  The chromatic numbers in these cases are (i) $\chi=2$ for
the line $T_n$; (ii) $\chi=2$ (3) for $C_n$ with $n$ even (odd); (iii) $\chi=3$
for the open triangular strip; and (iv) $\chi=3$ (4) for the cyclic triangular
strip with the number of triangles $N_t$ even (odd).  For the Potts
antiferromagnet with $nnn$ spin-spin couplings on the line (equivalently, on
the open triangular strip) with $n$ vertices, 
\beq
P(sq_d(L_y=1),FBC_y,FBC_x,q)=P(tri(L_y=2),FBC_y,FBC_x,q) = q(q-1)(q-2)^{n-2}
\label{psqdly1}
\eeq
and hence $W=q-2$, and $R_1$ is the full $q$ plane.  

For the Potts antiferromagnet with $nnn$ spin-spin couplings on the circuit,
the equivalence is with the cyclic triangular strip:
\beq
P(sq_d(L_y=1),FBC_y,PBC_x,q)=P(tri(L_y=2),FBC_y,PBC_x,q) \ . 
\label{psqdly1cyc}
\eeq
If the strip length involves an even number of triangles $N_t=n=2m$, then 
\cite{wcy,matmeth} 
\beq
P(tri(L_y=2),N_t=2m,FBC_y,PBC_x,q)=q^2-3q+1 + (q-2)^{2m} + (q-1)\Bigl [ 
(\lambda_{t2,3})^m + (\lambda_{t2,4})^m \Bigr ]
\label{ptrinteven}
\eeq
while for odd $N_t=n=2m+1$ \cite{t}
\beqs
& & P(tri(L_y=2),N_t=2m+1,FBC_y,PBC_x,q) = 
-(q^2-3q+1) + (q-2)[(q-2)^2]^m + \cr\cr
& & \frac{1}{2}(q-1)(q-3)\Biggl [\Bigl ( (\lambda_{t2,3})^m+ (\lambda_{t2,4})^m
\Bigr ) + \frac{\Bigl ( (\lambda_{t2,3})^m - (\lambda_{t2,4})^m \Bigr )}{
\lambda_{t2,3}-\lambda_{t2,4}} \Biggr ]
\label{ptrintodd}
\eeqs
where
\beq
\lambda_{t2,(3,4)} = \frac{1}{2}\biggl [ 5-2q \pm \sqrt{9-4q} \ \biggr ] \ .
\label{lamtly234}
\eeq
In both cases, $q_c=3$ and $W=q-2$ for $q \ge 3$ (and more generally, for $q$ 
in the region $R_1$ given in \cite{wcy}). 
Thus, the addition of next-nearest-neighbor couplings increases the value of 
$q$ beyond which there is nonzero ground state entropy from 2 to 3 and
decreases the value of the resultant entropy from $S_0=k_B\ln(q-1)$ to 
$S_0=k_B\ln(q-2)$ for $q \ge 3$. 

We proceed to consider the Potts antiferromagnet on the square
lattice, and again add next-nearest-neighbor couplings, or equivalently
redefine the lattice so that the bonds consists not just of the usual
horizontal and vertical bonds, but also of bonds connecting the diagonally
opposite vertices of each square.  Following our earlier notation \cite{w3}, we
shall denote this lattice as $sq_d$, where the $d$ refers to the addition of
these diagonal bonds.  For the square ($sq$) and $sq_d$ 
lattices, the chromatic numbers are 
\beq \chi(sq)=2 \ , \quad \chi(sq_d)=4 \ . 
\label{chi}
\eeq
No exact solution is known for $W(q)$ on the $sq_d$ lattice.  In the absence of
such an exact solution, Tsai and one of us (RS) have carried out Monte Carlo 
measurements of $W(sq_d,q)$ and have derived a rigorous lower bound \cite{w3}
\beq
W(sq_d,q) \ge \frac{(q-2)(q-3)}{q-1} \quad {\rm for} \quad q \ge \chi(sq_d) \
. 
\label{wsqdbound}
\eeq
This lower bound was compared with the actual value of $W(sq_d,q)$, as
determined by the Monte Carlo measurements for $5 \le q \le 10$ \cite{w3,ww} 
and was found 
to lie very close it (cf. Table III of \cite{w3}).  For example, for $q=6$ and
$q=8$, the ratio of the lower bound divided by the actual value was 0.981 and
0.995, and it increased monotonically toward unity as $q$ increased. 
Since it is possible to obtain exact analytic solutions for $W$ on
infinite-length, finite-width strips of 2D lattices \cite{bds},\cite{w},
\cite{strip}-\cite{w2d},\cite{pg,wcy,pm,tk,bcc,t}, one 
has an alternate way to investigate $W(sq_d,q)$, namely to calculate $W$ 
exactly on strips of the $sq_d$ lattice, with various boundary conditions.  
It 
has, indeed, been found \cite{w2d} that for the square and triangular lattices,
the values of $W$ for such infinite-length strips of even rather modest widths
are close to the corresponding values for the 2D thermodynamic limit, for
moderate values of $q$.  In the present work we report exact calculations of 
$P(q)$, $W(q)$, and ${\cal B}$ on strips of the $sq_d$ lattice with various 
boundary conditions.  The longitudinal and transverse directions on the strip 
are taken to be $\hat x$ and $\hat y$, respectively.  In Fig. 1 we show some
illustrative strips of the $sq_d$ lattice. 

\vspace{6mm}

\unitlength 1.3mm
\hspace*{5cm}
\begin{picture}(40,10)
\multiput(0,0)(10,0){5}{\circle*{2}}
\multiput(0,10)(10,0){5}{\circle*{2}}
\multiput(0,0)(10,0){5}{\line(0,1){10}}
\multiput(0,0)(0,10){2}{\line(1,0){40}}
\multiput(0,0)(10,0){4}{\line(1,1){10}}
\multiput(0,10)(10,0){4}{\line(1,-1){10}}
\put(-2,-2){\makebox(0,0){5}}
\put(8,-2){\makebox(0,0){6}}
\put(18,-2){\makebox(0,0){7}}
\put(28,-2){\makebox(0,0){8}}
\put(38,-2){\makebox(0,0){5}}
\put(-2,12){\makebox(0,0){1}}
\put(8,12){\makebox(0,0){2}}
\put(18,12){\makebox(0,0){3}}
\put(28,12){\makebox(0,0){4}}
\put(38,12){\makebox(0,0){1}}
\put(20,-8){\makebox(0,0){(a)}}
\end{picture}
\vspace*{2cm}

\hspace*{5cm}
\begin{picture}(40,20)
\multiput(0,0)(10,0){5}{\circle*{2}}
\multiput(0,10)(10,0){5}{\circle*{2}}
\multiput(0,20)(10,0){5}{\circle*{2}}
\multiput(0,0)(10,0){5}{\line(0,1){20}}
\multiput(0,0)(0,10){3}{\line(1,0){40}}
\put(0,10){\line(1,1){10}}
\multiput(0,0)(10,0){3}{\line(1,1){20}}
\put(30,0){\line(1,1){10}}
\put(0,10){\line(1,-1){10}}
\multiput(0,20)(10,0){3}{\line(1,-1){20}}
\put(30,20){\line(1,-1){10}}
\put(-2,-2){\makebox(0,0){9}}
\put(8,-2){\makebox(0,0){10}}
\put(18,-2){\makebox(0,0){11}}
\put(28,-2){\makebox(0,0){12}}
\put(38,-2){\makebox(0,0){9}}
\put(-2,12){\makebox(0,0){5}}
\put(8,12){\makebox(0,0){6}}
\put(18,12){\makebox(0,0){7}}
\put(28,12){\makebox(0,0){8}}
\put(38,12){\makebox(0,0){5}}
\put(-2,22){\makebox(0,0){1}}
\put(8,22){\makebox(0,0){2}}
\put(18,22){\makebox(0,0){3}}
\put(28,22){\makebox(0,0){4}}
\put(38,22){\makebox(0,0){1}}
\put(20,-8){\makebox(0,0){(b)}}
\end{picture}
\vspace*{2cm}

\hspace*{5cm}
\begin{picture}(40,30)
\multiput(0,0)(10,0){5}{\circle*{2}}
\multiput(0,10)(10,0){5}{\circle*{2}}
\multiput(0,20)(10,0){5}{\circle*{2}}
\multiput(0,30)(10,0){5}{\circle*{2}}
\multiput(0,0)(10,0){5}{\line(0,1){30}}
\multiput(0,0)(0,10){4}{\line(1,0){40}}
\put(0,20){\line(1,1){10}}
\put(0,10){\line(1,1){20}}
\multiput(0,0)(10,0){2}{\line(1,1){30}}
\put(20,0){\line(1,1){20}}
\put(30,0){\line(1,1){10}}
\put(0,10){\line(1,-1){10}}
\put(0,20){\line(1,-1){20}}
\multiput(0,30)(10,0){2}{\line(1,-1){30}}
\put(20,30){\line(1,-1){20}}
\put(30,30){\line(1,-1){10}}
\put(-2,-2){\makebox(0,0){1}}
\put(8,-2){\makebox(0,0){2}}
\put(18,-2){\makebox(0,0){3}}
\put(28,-2){\makebox(0,0){4}}
\put(38,-2){\makebox(0,0){1}}
\put(-2,12){\makebox(0,0){9}}
\put(8,12){\makebox(0,0){10}}
\put(18,12){\makebox(0,0){11}}
\put(28,12){\makebox(0,0){12}}
\put(38,12){\makebox(0,0){9}}
\put(-2,22){\makebox(0,0){5}}
\put(8,22){\makebox(0,0){6}}
\put(18,22){\makebox(0,0){7}}
\put(28,22){\makebox(0,0){8}}
\put(38,22){\makebox(0,0){5}}
\put(-2,32){\makebox(0,0){1}}
\put(8,32){\makebox(0,0){2}}
\put(18,32){\makebox(0,0){3}}
\put(28,32){\makebox(0,0){4}}
\put(38,32){\makebox(0,0){1}}
\put(20,-8){\makebox(0,0){(c)}}
\end{picture}
\vspace*{2cm}

\begin{figure}[h]
\caption{\footnotesize{Illustrative strip graphs of the $sq_d$ lattice: (a,b),
$L_y=2,3$ $(FBC_y,PBC_x)$ (cyclic); (c) $L_y=3$, $(PBC_y,PBC_x)$ (toroidal
boundary conditions).}}
\label{fig1}
\end{figure}

An important property is that with the two added diagonal bonds, each square of
the $sq_d$ lattice constitutes a complete graph on four vertices.  (Here, the 
complete graph on $r$ vertices, $K_r$, is defined as the graph each of whose 
vertices is connected to all of the other $r-1$ vertices by bonds (= edges);
it has chromatic number $\chi(K_r)=r$.)  Compared with the square lattice,
for a given $q$-coloring of the $sq_d$ lattice, the addition of these bonds
clearly increases the constraints on the coloring of each vertex and therefore
decreases $P(G,q)$.  As we proved earlier \cite{wn}, if a lattice
$\Lambda^\prime$ can be obtained from another, $\Lambda$, by connecting
disjoint vertices of $\Lambda$ with bonds, then $W(\Lambda^\prime,q) \le
W(\Lambda,q)$ for $q$-colorings of the two lattices.  An example of the
application of this theorem was given in \cite{wn}: for $q$ colorings of
the square, triangular (tri), and honeycomb (hc) lattices, $W(tri,q)
< W(sq,q) \le W(hc,q)$.  ($W(sq,q)$ is strictly less than $W(hc,q)$ except at
the value $q=2$, where $W(sq,2)=W(hc,2)=1$.) In the present context, we note 
the inequality for $q$-colorings of these lattices: 
\beq 
W(sq_d,q) < W(tri,q) < W(sq,q) \le W(hc,q)
\label{wineqality}
\eeq
(for $q$ values where such colorings are possible). 

We use the symbols FBC$_y$ and PBC$_y$ for free and periodic transverse
boundary conditions and, as above, FBC$_x$, PBC$_x$, and TPBC$_x$ for free,
periodic, and twisted periodic longitudinal boundary conditions.  The term
``twisted'' means that the longitudinal ends of the strip are identified with
reversed orientation.  These strip graphs can be embedded on surfaces with the
following topologies: (i) (FBC$_y$,FBC$_x$): open strip; (ii)
(PBC$_y$,FBC$_x$): cylindrical; (iii) (FBC$_y$,PBC$_x$): cylindrical (denoted
cyclic here); (iv) (FBC$_y$,TPBC$_x$): M\"obius; (v) (PBC$_y$,PBC$_x$): torus;
and (vi) (PBC$_y$,TPBC$_x$): Klein bottle.\footnote{\footnotesize{These BC's
can all be implemented in a manner that is uniform in the length $L_x$; the
case (vii) (TPBC$_y$,TPBC$_x$) with the topology of the projective plane
requires different identifications as $L_x$ varies and will not be considered
here.}}

The labelling of the strips generally follows our earlier labelling
conventions.  Thus for a strip with free transverse and longitudinal boundary
conditions, $(FBC_y,FBC_x)$, the length of the strip is taken to be $m+1$
squares or equivalently edges, with $L_x=m+2$, and the width is $L_y$ vertices.
This strip thus has $n=L_xL_y$ vertices and $e=4L_xL_y-3(L_x+L_y)+2$ edges.
For cyclic strips, the width is defined in the same manner and the length is
$L_x$ vertices or equivalently edges.  For strips with periodic transverse
boundary conditions, including $(PBC_y,FBC_x)$ and $(PBC_y,PBC_x)$, a width of
$L_y=3$ means that the cross section involves $L_y$ vertices.  For $L_x=m \ge
3$ to avoid certain degenerate cases, the $sq_d$ strips with either cyclic or
torus boundary conditions have $n=L_xL_y$ vertices. With the same restriction,
the cyclic strips have $e=L_x(4L_y-3)$ edges and the torus strips have
$e=4L_xL_y$ edges.

Let us next comment on the planarity or nonplanarity of the strips of the
$sq_d$ lattice with various boundary conditions.  We shall concentrate here on
nondegenerate cases where the strips are proper graphs without multiple 
edges.  Consider first the strips
with $(FBC_y,FBC_x)$ boundary conditions.  For $L_y=2$ and arbitrary $L_x$, it
is easy to show that these are planar by taking the second diagonal bond for
each square and drawing it external to the strip; by redefining the labelling
of the $x$ and $y$ axes, it follows that this strip with $(FBC_y,FBC_x)$
boundary conditions, $L_x=2$, and arbitrary $L_y$ is also planar.  For other
cases we shall make use of two theorems from graph theory.  The first of these
states that if $G$ is a planar graph with $n$ vertices and $e$ edges with $n
\ge 3$, then $e \le 3(n-2)$ (e.g. Corollary 11.1(c) in \cite{harary}) and the
second states that if $G$ is a planar graph with $n \ge 4$, then $G$ has at
least four vertices of degree $\Delta \le 5$ (e.g.  Corollary 11.1(e) in
\cite{harary}).  Now 
\beq
3(n-2)-e = -L_xL_y +3(L_x+L_y) -8 \quad {\rm for} \quad sq_d, \ \ (FBC_y,FBC_x)
\label{etestopen}
\eeq
so that for sufficiently great $L_x$ and/or $L_y$, $3(n-2)-e$ is negative and
hence the strip is nonplanar, by the first theorem.  For example, $3(n-2)-e <
0$ if $L_y=4$ and $L_x \ge 5$ or vice versa, i.e., $L_x=4$ and $L_y \ge 5$; and
similarly, if $L_x=L_y \ge 5$.  Considering next the strips of the $sq_d$ 
lattice that are cyclic, i.e., have $(FBC_y,PBC_x)$ boundary conditions, we
have
\beq
3(n-2)-e = -L_xL_y+3L_x-6 \quad {\rm for} \quad sq_d, \ \ (FBC_y,PBC_x) \ . 
\label{etestcyclic}
\eeq
Hence, $3(n-2)-e < 0$ for all $L_x$ if $L_y \ge 3$, so that these strips are 
nonplanar.  For the strips with $(PBC_y,PBC_x)$, i.e., torus boundary 
conditions, we have
\beq
3(n-2)-e = -(L_xL_y+6) \quad {\rm for} \quad sq_d, \ \ (PBC_y,PBC_x)
\label{etesttorus}
\eeq
so that these strips are also nonplanar.  This can be seen alternatively by
observing that each vertex on the torus strips has degree $\Delta=8$ and
applying the second theorem cited above.  The second theorem also shows that
the $sq_d$ strip with Klein bottle boundary conditions is nonplanar.

A generic form for chromatic polynomials for recursively defined families of
graphs, of which strip graphs $G_s$ are special cases, is
\beq
P((G_s)_m,q) =  \sum_{j=1}^{N_{G_s,\lambda}} c_{G_s,j}(q)(\lambda_{G_s,j}(q))^m
\label{pgsum}
\eeq
where $c_{G_s,j}(q)$ and the $N_{G_s,\lambda}$ terms $\lambda_{G_s,j}(q)$ 
depend on the type of strip graph $G_s$, as indicated, but are independent of 
$m$.

\section{Strips with $(FBC_{\lowercase{y}},FBC_{\lowercase{x}})$}

\subsection{$L_y=2$}

The chromatic polynomial for the strip of the $sq_d$ lattice with $L_y=1$ and
free transverse and longitudinal boundary conditions was given above in 
(\ref{psqdly1}).  For the $L_y=2$ case the chromatic polynomial is
\beq
P(sq_d(L_y=2)_m,FBC_y,FBC_x,q) = q(q-1)\Bigl [ (q-2)(q-3) \Bigr ]^{m+1} \ . 
\label{pk4ff}
\eeq
In the $m \to \infty$ limit, 
\beq
W(sq_d(L_y=2),FBC_y,FBC_x,q)=\Bigl [ (q-2)(q-3) \Bigr ]^{1/2}
\label{wk4ff}
\eeq
with ${\cal B}=\emptyset$. 

\subsection{$L_y=3$} 

For the $L_y=3$ strip, we use the same generating function method as we have
before \cite{strip,hs,wcy}.  In general, for the family of strip graphs $G_s$,
the generating function $\Gamma(G_s,q,x)$ is a rational function of the form
\beq
\Gamma(G_s,q,x) = \frac{{\cal N}(G_s,q,x)}{{\cal D}(G_s,q,x)}
\label{gammagen}
\eeq
with
\beq
{\cal N}(G_s,q,x) = \sum_{j=0}^{d_{\cal N}} A_{G_s,j}(q) x^j
\label{n}
\eeq
and
\beq
{\cal D}(G_s,q,x) = 1 + \sum_{j=1}^{d_{\cal D}} b_{G_s,j}(q) x^j
\label{d}
\eeq
where the $A_{G_s,i}$ and $b_{G_s,i}$ are polynomials in $q$ (with no common
factors) and
\beq
d_{\cal N} = deg_x({\cal N})
\label{dn}
\eeq
and
\beq
d_{\cal D} = deg_x({\cal D}) \ .
\label{dd}
\eeq
This generating function yields the chromatic polynomials via the Taylor series
expansion in the auxiliary variable $x$: 
\beq
\Gamma(G_s,q,x) = \sum_{m=0}^{\infty}P((G_s)_m,q)x^m
\label{gamma}
\eeq
where we follow the notational convention in \cite{strip}, according to
which a strip is considered to be comprised of $m$ repetitions of a basic
subgraph unit $H$ connected to an initial subgraph $I$; here we take $I=H$ so
that a strip with a given value of $m$ has $m+1$ columns of $K_4$'s and $m+2$
vertices in the longitudinal direction.  
The denominator can be written in factorized form as 
\beq
{\cal D}_{G_s} = \prod_{j=1}^{d_{\cal D}}(1-\lambda_{G_s,j} x) \ . 
\label{denfac}
\eeq
These are the $\lambda_{G_s,j}$'s in eq. (\ref{pgsum}); the coefficients are
determined by eqs. (2.14) or (2.19) in \cite{hs}. 

For $L_y=3$, we find $d_{\cal D}=2$, $d_{\cal N}=1$, and 
\beq
\Gamma(sq_d(L_y=3),FBC_y,FBC_x,q,x)=\frac{q(q-1)(q-2)(q-3)^2\Bigl [ 
(q-2) -(q-1)(q-3)x \Bigr ]}{1-(q-3)(q^2-6q+11)x + (q-2)(q-3)^3x^2} \ . 
\label{gammak4ly3ff}
\eeq
The denominator can be written as 
\beq
{\cal D}_{sqd3o} = (1-\lambda_{sqd3o,1} x)(1-\lambda_{sqd3o,2} x)
\label{dk4ly3ff}
\eeq
where
\beq
\lambda_{sqd3o,(1,2)} = \frac{1}{2}(q-3)\biggl [ q^2-6q+11 \pm \Bigl (
q^4-12q^3+54q^2-112q+97 \Bigr )^{1/2} \biggr ]
\label{lamk412}
\eeq
and the shorthand $sqd3o$ denotes the strip of the $sq_d$ lattice with $L_y=3$
and open (o) boundary conditions. 
 From this, using the general formulas in \cite{hs}, one can write the 
chromatic polynomial in the form of eq. (\ref{pgsum}) (with $N_\lambda=2$).  
In the $m \to \infty$ limit, 
\beq
W(sq_d(L_y=3),FBC_y,FBC_x,q) = (\lambda_{sqd3o,1})^{1/3} \ . 
\label{wk4ly3ff}
\eeq

\begin{figure}[[hbtp]
\centering
\leavevmode
\epsfxsize=4.0in
\begin{center}
\leavevmode
\epsffile{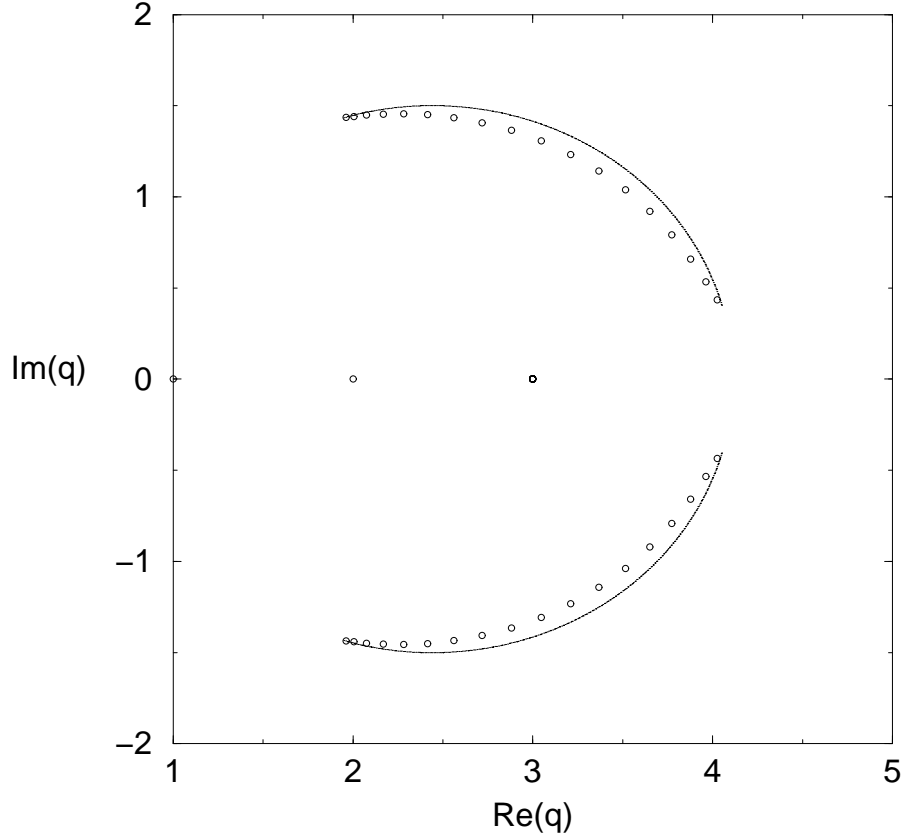}
\end{center}
\caption{\footnotesize{Locus ${\cal B}$ for $W$ for the $3 \times \infty$
strip of the $sq_d$ lattice with free transverse and longitudinal boundary
conditions. Chromatic zeros are shown for the case $L_x=20$ (i.e., $n=60$).}}
\label{k4y3}
\end{figure}

The nonanalytic locus ${\cal B}$ is shown in Fig. \ref{k4y3} and is comprised
of an arc stretching between endpoints at $q \simeq 1.95 + 1.43i$ and $4.05 +
0.396i$, together with the complex conjugate arc.  In agreement with the
general discussion given before \cite{strip,hs,bcc}, these four points are the
branch points of the square root in eq. (\ref{lamk412}).  ${\cal B}$ does not
intersect the real $q$ axis, so that no $q_c$ is defined.  The region $R_1$ is
the entire $q$ plane, with the exception of the arcs lying on ${\cal B}$.

\subsection{$L_y=4$}

Here we find $d_{\cal D}=4$, $d_{\cal N}=3$.  Again using the shorthand
notation $sqd4o$ to denote this open strip of the $sq_d$ lattice with $L_y=4$
we have 
\beq
b_{sqd4o,1}=-q^4+13q^3-68q^2+171q-176
\label{bsqd4o_1}
\eeq
\beq
b_{sqd4o,2}=(q-3)(2q^5-33q^4+219q^3-729q^2+1214q-803)
\label{bsqd4o_2}
\eeq
\beq
b_{sqd4o,3}=(q-3)^3(q^5-17q^4+118q^3-420q^2+770q-586)
\label{bsqd4o_3}
\eeq
\beq
b_{sqd4o,4}=-(q-2)(q-3)^6(q-4) \ . 
\label{bsqd4o_4}
\eeq
For the functions $A_{sqd4,o,j}$ in the numerator, ${\cal N}$, it is convenient
to extract a common factor and thus define 
\beq
A_{sqd4o,j}=q(q-1)(q-2)(q-3)^3 \bar A_{sqd4o,j} \ . 
\label{asqd4factor}
\eeq
Then
\beq
\bar A_{sqd4o,0}=(q-2)^2
\label{asqd4o_0}
\eeq
\beq
\bar A_{sqd4o,1}=-(2q^4-21q^3+78q^2-113q+47)
\label{asqd4o_1}
\eeq
\beq
\bar A_{sqd4o,2}=-(q-3)(q^5-14q^4+77q^3-204q^2+245q-91)
\label{asqd4o_2}
\eeq
and
\beq
\bar A_{sqd4o,3}=(q-1)^2(q-3)^3(q-4) \ . 
\label{asqd4o_3}
\eeq
Let us write the denominator as 
\beq
{\cal D}_{sqd4o} = \prod_{j=1}^4 (1-\lambda_{sqd4o,j} x) \ . 
\label{dk4ly3odenfac}
\eeq
Then
\beq
W=(\lambda_{sqd4o,j,max})^{1/4} \quad {\rm for} \quad q \in R_1
\label{wr14open}
\eeq
where $\lambda_{sqd4o,j,max}$ is the $\lambda_{sqd4o,j}$ in
(\ref{dk4ly3odenfac}) with maximal magnitude in this region. 

\begin{figure}[[hbtp]
\centering
\leavevmode
\epsfxsize=4.0in
\begin{center}
\leavevmode
\epsffile{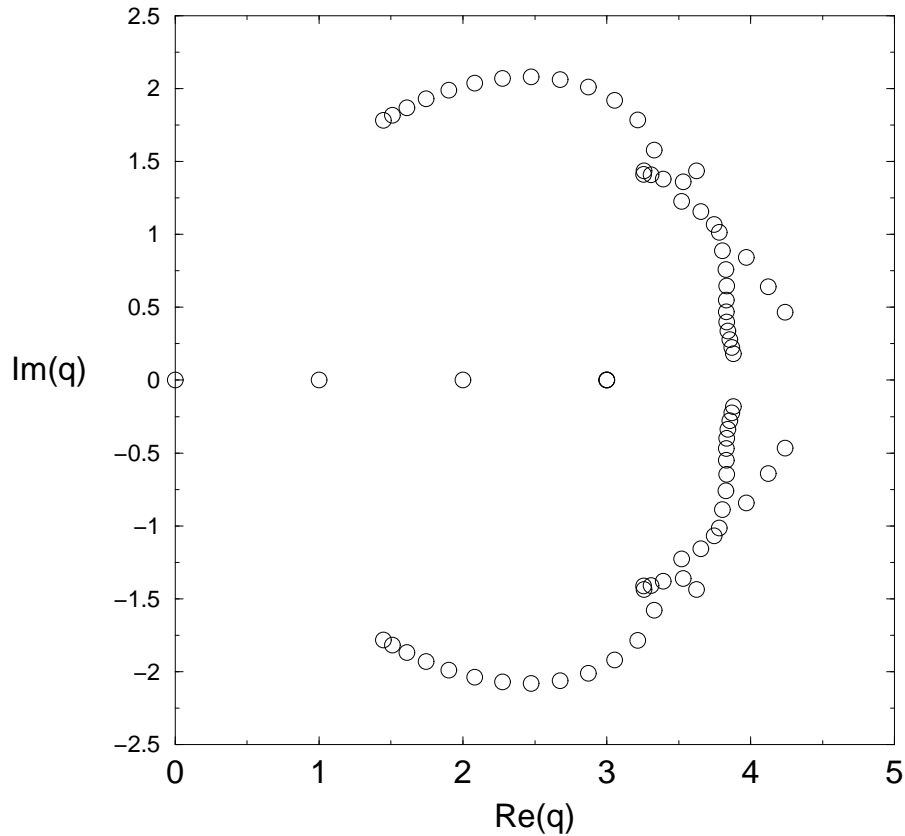}
\end{center}
\caption{\footnotesize{Chromatic zeros for the $4 \times L_x$ 
strip of the $sq_d$ lattice with free boundary conditions and $L_x=20$
(i.e., $n=80$).}}
\label{k4y4}
\end{figure}

Chromatic zeros are shown in Fig. \ref{k4y4} for $L_x=20$, i.e., $n=80$.  For
this great a length, these chromatic zeros give a reasonably good approximation
to the asymptotic locus ${\cal B}$.  As is evident from this figure, ${\cal B}$
does not cross the real $q$ axis, so that no $q_c$ is defined.  

\section{Strips with $(FBC_{\lowercase{y}},(T)PBC_{\lowercase{x}})$}

\subsection{$L_y=2$}

The chromatic polynomial for the $L_y=1$ cyclic strip of the $sq_d$ lattice was
given above in eqs. (\ref{ptrinteven}) and (\ref{ptrintodd}). 
Here we consider the $L_y=2$ strip of the $sq_d$ lattice with $(FBC_y,PBC_x)$,
i.e. cyclic, boundary conditions.  For a given value of $L_x$, this cyclic
strip graph is identical to the corresponding strip with M\"obius boundary
conditions $(FBC_y,TPBC_x)$:
\beq
G(sq_d,L_y=2,L_x,FBC_y,PBC_x) = G(sq_d,L_y=2,L_x,FBC_y,TPBC_x)
\label{cyceqmobly2}
\eeq
This can be proved by calculating the adjacency matrices for the cyclic and
M\"obius strips, which are identical.  (Here the adjacency matrix of an
$n$-vertex graph is the $n \times n$ matrix $A$ with $A_{ij}$ equal to the
number of bonds (edges) that connect the $i$'th and $j$'th vertices.  The
adjacency matrix fully defines the graph.)  Because of the identity of the
$L_y=2$ cyclic and M\"obius strips, we shall refer to them both with the
designation $sq_d(L_y=2)_m,FBC_y,(T)PBC_x)$.  For the general cyclic strip of
the $sq_d$ strip, with $L_x \ge 4$ to avoid degenerate cases, the chromatic 
number is given by 
\beq
\chi(sq_d,L_y,L_x,FBC_y,PBC_x)= \cases{ 4 & if $L_x$ is even \cr
                                        5 & if $L_x$ is odd \cr }
\label{chivalscyc}
\eeq
For the present $L_y=2$ cyclic/M\"obius strips, the degenerate cases 
are as follows: for $L_x=2$, the strip reduces to $K_4$ while for $L_x=3$ it 
reduces to $K_6$, with $\chi(K_p)=p$.

The chromatic polynomial for the cyclic strip of the $sq_d$ lattice with 
$L_y=2$ and $L_x \equiv m$ is \cite{readcarib}
\beq
P(sq_d(L_y=2),FBC_y,(T)PBC_x)_m,q)= \frac{1}{2}q(q-3)2^m+[(q-2)(q-3)]^m+
(q-1)[2(3-q)]^m \ .
\label{pk4ly2cyc}
\eeq
We determine the boundary ${\cal B}$ to be the union of a circle centered at 
$q=2$ with radius 2 and a circle centered at $q=3$ with radius 1:
\beq
{\cal B}: \{|q-2|=2 \} \ \cup \ \{|q-3|=1 \} \ . 
\label{bk4ly2cyc}
\eeq
These two circles osculate (i.e., intersect with equal tangents) at $q_c$,
where 
\beq
q_c(sq_d(L_y=2),FBC_y,(T)PBC_x)=4 \ . 
\label{qcly2}
\eeq
In the terminology of algebraic geometry, this point $q_c$ is thus a tacnode. 

\begin{figure}[[hbtp]
\centering
\leavevmode
\epsfxsize=4.0in
\begin{center}
\leavevmode
\epsffile{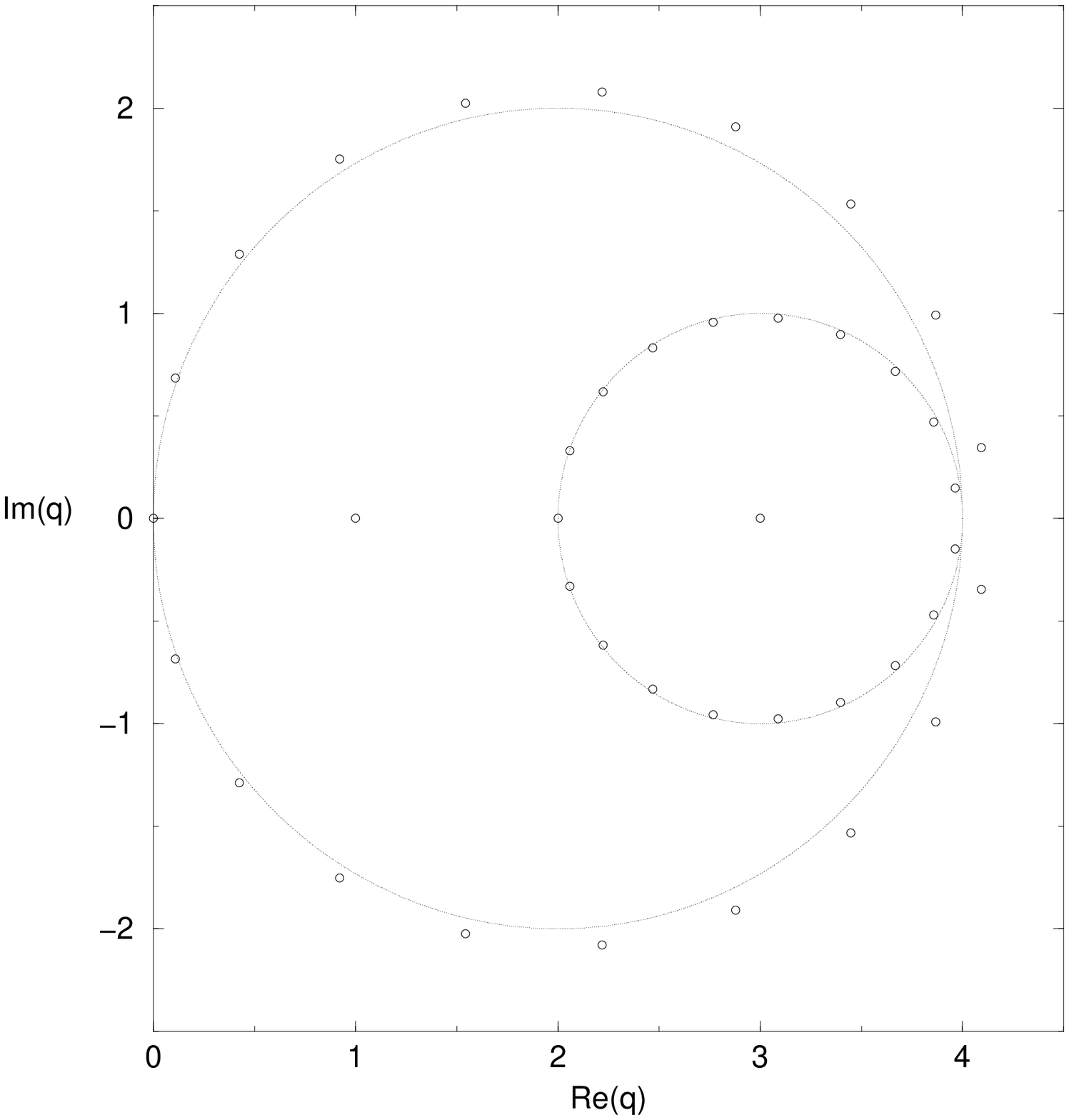}
\end{center}
\caption{\footnotesize{Locus ${\cal B}$ for $W$ for the $2 \times \infty$
cyclic or M\"obius strip of the $sq_d$ lattice.  Chromatic zeros are shown for
the case $L_x=20$ (i.e., $n=40$).}}
\label{k4pxy2}
\end{figure}

This locus is shown in Fig. \ref{k4pxy2}.  The locus ${\cal B}$ separates the
$q$ plane into three regions: (i) the outermost region $R_1$ which is the
exterior of the larger circle, $|q-2| \ge 2$ and which thus includes the real
intervals $q \ge 4$ and $q \le 0$; (ii) region $R_2$ which is the interior of
the smaller circle, $|q-3| \le 1$ and includes the real interval $2 \le q \le
4$; and (iii) region $R_3$ which is the interior of the larger circle $|q-2|
\le 2$ minus the smaller disk $|q-3|=1$ and includes the real interval $0 \le q
\le 2$.  Thus, ${\cal B}$ crosses the real $q$ axis at $q=0,2,4$ and $q_c=4$.
As is evident in Fig. \ref{k4pxy2}, the chromatic zeros lie near to the
asymptotic locus ${\cal B}$. In the various regions 
\beq 
W=[(q-2)(q-3)]^{1/2} \quad {\rm for} \quad q \in R_1
\label{wk4ly2cycr1}
\eeq
\beq
|W|=2^{1/2} \quad {\rm for} \quad q \in R_2
\label{wk4ly2cycr2}
\eeq
\beq
|W|=|2(q-3)|^{1/2} \quad {\rm for} \quad q \in R_3
\label{wk4ly2cycr3}
\eeq
(for $q$ in regions other than $R_1$, only the magnitude $|W(q)|$ can be
determined unambiguously).

We define the sum of the coefficients as 
\beq
C(G)=\sum_{j=1}^{N_{\lambda_G}} c_{G,j} \ .
\label{cgsum}
\eeq
For sufficiently large positive integer $q$, the coefficient
$c_{G,j}$ in (\ref{cgsum}) can be interpreted as the multiplicity of the
corresponding eigenvalue $\lambda_{G,j}$ of the coloring matrix, i.e., the
dimension of the corresponding invariant subspace in the full space of coloring
configurations \cite{b,ww,matmeth}.  We recall that the coloring matrix can be
defined as the matrix whose $i,j$ element is 1 (0) if the coloring
configurations on two adjacent transverse slices of the strip are compatible
(incompatible).  Thus, in the absence of zero eigenvalues of the coloring
matrix, $C(G)$ is the dimension of the space of coloring configurations of such
a transverse slice.  For the cyclic strip graphs of the $sq_d$ lattice, the
transverse slice is the line graph $L_n$ of length $L_y$ vertices, so the space
of coloring configurations of this transverse slice is
\beq
P(L_n,q)=P(T_n,q)=q(q-1)^n \ . 
\label{ptree}
\eeq
In cases where the cyclic strip and the M\"obius strip of the $sq_d$ lattice
are identical, the full coloring matrix automatically takes account of both
contributions, so that for each individual strip, corresponding to a given
permutation in the identification of vertices at the longitudinal boundary, one
must divide by the symmetry factor.  In the present case, $L_y=2$ and there are
two permutations of the identifications of the boundary conditions that give
identical strip graphs, so that this symmetry factor is $1/2!$ so that
\beq
C(sq_d,L_y=2,FBC_y,PBC_x) = C(sq_d,L_y=2,FBC_y,TPBC_x)=\frac{1}{2}q(q-1)
\label{csumcycly2}
\eeq
This agrees with the sum of the coefficients in the expression
(\ref{pk4ly2cyc}).  A remark that will be relevant later is that 
if the coloring matrix has a zero
eigenvalue of multiplicity $c_{zero}$, then, since this eigenvalue does not
appear in (\ref{pgsum}), the sum of the coefficients that do appear in the
chromatic polynomial (\ref{pgsum}) is equal to the full dimension of the space
of coloring configurations minus $c_{zero}$.  

\subsection{$L_y=3$}

We have calculated the chromatic polynomial for the next wider cyclic strip, 
with $L_y=3$.  For this strip the chromatic numbers $\chi$ are the
same as for the $L_y=2$ $sq_d$ strip.  
We find $N_{sqd,L_y=3,cyc.,\lambda}=16$ and 
\beq
P(sq_d(L_y=3),FBC_y,PBC_x)_m,q)=\sum_{j=1}^{16} c_{sqd3c,j}
(\lambda_{sqd3c,j})^m
\label{pk4ly3cyc}
\eeq
(The terms $\lambda_{sqd3c,j}$ are the same for cyclic and M\"obius
longitudinal boundary conditions.)  With the $\lambda_{sqd3c,j}$'s 
ordered according to decreasing degrees of their coefficients for the cyclic
strip, we find 
\beq
\lambda_{sqd3c,1}=1
\label{lamsqd3c1}
\eeq
\beq
\lambda_{sqd3c,2}=-2
\label{lamsqd3c2}
\eeq
\beq
\lambda_{sqd3c,3}=-3
\label{lamsqd3c3}
\eeq
\beq
\lambda_{sqd3c,4}=q-3
\label{lamsqd3c4}
\eeq
\beq
\lambda_{sqd3c,5}=-(q-3)
\label{lamsqd3c5}
\eeq
and
\beq
\lambda_{sqd3c,(6,7)}=q-4 \pm \sqrt{2q^2-14q+25} \ . 
\label{lamsqd3d67}
\eeq 
The terms $\lambda_{sqd3c,j}$, $j=8,9,10$ are the roots of the cubic
equation 
\beq \xi^3-4(q-4)\xi^2+(q^2-10q+17)\xi + 2(q-1)(q-3) \ . 
\label{eqcub1}
\eeq
For $j=11$ we have
\beq
\lambda_{sqd3c,11}=-(q-3)^2 \ . 
\label{lamsqd311}
\eeq
The terms $\lambda_{sqd3c,j}$, $j=12,13,14$ are the roots of the cubic 
equation
\beq
\xi^3+(2q^2-15q+30)\xi^2-(q-3)^2(q^2-5q+5)\xi-2(q-3)^4 \ . 
\label{eqcub2}
\eeq
Finally, 
\beq
\lambda_{sqd3c,(15,16)}=\lambda_{sqd3o,(1,2)} \ . 
\label{lamsqd31516}
\eeq
For the coefficients we calculate 
\beq
c_{sqd3c,1}=\frac{1}{6}(q-1)(q-2)(q-3)
\label{csqd3_1}
\eeq
\beq
c_{sqd3c,2}=\frac{1}{3}q(q-2)(q-4)
\label{csqd3_2}
\eeq
\beq
c_{sqd3c,3}=\frac{1}{6}q(q-1)(q-5)
\label{csqd3_3}
\eeq
\beq
c_{sqd3c,j}=\frac{1}{2}q(q-3)  \quad {\rm for} \quad j=4,8,9,10
\label{csqd3_4}
\eeq
\beq
c_{sqd3c,j}=\frac{1}{2}(q-1)(q-2) \quad {\rm for} \quad j=5,6,7
\label{csqd3_567}
\eeq
\beq
c_{sqd3c,j}=q-1 \quad {\rm for} \quad j=11,12,13,14
\label{csqd3_1114}
\eeq
and
\beq
c_{sqd3c,j}=1 \quad {\rm for} \quad j=15,16  \ . 
\label{csqd3_1516}
\eeq
Summing the coefficients $c_{sqd3c,j}$, we find 
\beq
C(sq_d(L_y=3),FBC_y,PBC_x)=\frac{1}{6}q(q+1)(4q-7) \ . 
\label{csum3}
\eeq
Since for $L_y=3$ the cyclic and M\"obius strips of the $sq_d$ lattice are
distinct, the full sum of eigenvalue multiplicities is equal to (\ref{ptree})
with $n=L_y=3$ (without dividing by any symmetry factor). The sum of the 
coefficients appearing in (\ref{pk4ly3cyc}) is less than this quantity, 
$q(q-1)^2$, by the amount 
\beq
c_{sqd3c,zero}=\frac{1}{6}q(2q^2-9q+13)
\label{csqd3czero}
\eeq
which indicates that for this strip the coloring matrix has a zero eigenvalue
with this multiplicity.  

\begin{figure}[p]
\centering
\leavevmode
\epsfxsize=4.0in
\begin{center}
\leavevmode
\epsffile{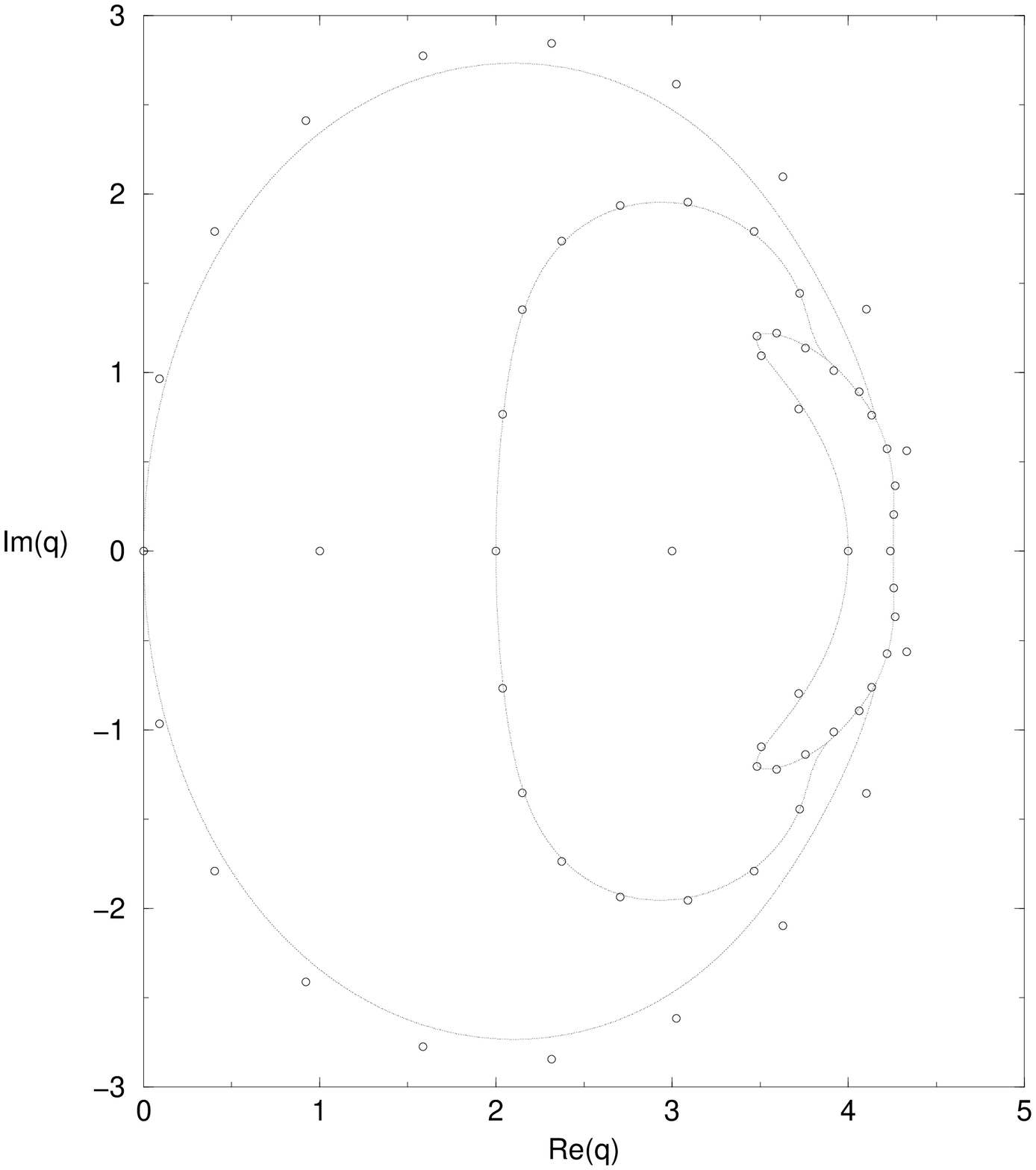}
\end{center}
\caption{\footnotesize{Locus ${\cal B}$ for $W$ for the $3 \times \infty$
cyclic or M\"obius strip of the $sq_d$ lattice.  Chromatic zeros are shown for
the case $L_x=20$ (i.e., $n=60$).}}
\label{k4pxy3}
\end{figure}

The boundary ${\cal B}$ is shown in Fig. \ref{k4pxy3}.  This boundary separates
the $q$ plane into four regions.  These include (i) the outermost region 
$R_1$, which contains the semi-infinite intervals $q > q_c$ and $q < 0$, where 
$q_c$ is
\beq
q_c(sq_d(L_y=3),FBC_y,PBC_x)=4.254654...
\label{qcsqd3c}
\eeq
(a root of the equation $q^4-14q^3+72q^2-168q+162=0$); (ii) a narrow
crescent-shaped region $R_2$ containing the real interval $4 \le q \le q_c$; 
(iii) the region $R_3$ containing the real interval $2 \le q \le 4$; and 
(iv) the region $R_4$ containing the real interval $0 \le q \le 2$.  
Associated with these regions are two complex-conjugate pairs of triple points,
as is evident in Fig. \ref{k4pxy3}. Note that $q_c$ is not a
tacnode for the ($L_x \to \infty$ limit of the) $L_y=3$ cyclic strip, in 
contrast to the situation for the corresponding $L_y=2$ cyclic strip. 

In the various regions
\beq
W=(\lambda_{sqd3c,15})^{1/3} \quad {\rm for} \quad q \in R_1
\label{wsqd3cr1}
\eeq
\beq
|W|=3^{1/3} \quad {\rm for} \quad q \in R_2
\label{wsqd3cr2}
\eeq
and 
\beq
|W|=|\lambda_{sqd3c,810m}|^{1/3} \quad {\rm for} \quad q \in R_3
\label{wsqd3cr3}
\eeq
where $\lambda_{sqd3c,810m}$ denotes the root of the cubic (\ref{eqcub1}) of
maximal magnitude in $R_3$, and 
\beq
|W|=|\lambda_{sqd3c,1214m}|^{1/3} \quad {\rm for} \quad q \in R_4
\label{wsqd3cr4}
\eeq
where $\lambda_{sqd3c,1214m}$ denotes the root of the cubic (\ref{eqcub2}) of
maximal magnitude in $R_4$. 

\section{Strips with $(PBC_{\lowercase{y}},FBC_{\lowercase{x}})$}

\subsection{$L_y=3$}

For the $L_y=3$, $L_x=m+2$ strip of $K_4$'s forming squares, with
(PBC$_y$,FBC$_x$) boundary conditions (for which $n=3(m+2)$) we find
\beq
P(sq_d(L_y=3)_m,PBC_y,FBC_x,q) = q(q-1)(q-2)\Bigl [ (q-3)(q-4)(q-5) \Bigr
]^{m+1} 
\label{ppbcy}
\eeq
whence 
\beq
W(sq_d(L_y=3),PBC_y,FBC_x,q) = \Bigl [ (q-3)(q-4)(q-5)\Bigr ]^{1/3} \ . 
\label{wk4ly3pbcyfbcx}
\eeq
The continuous nonanalytic locus ${\cal B}=\emptyset$; $W$ has isolated 
branch point singularities where it vanishes at $q=3,4$, and 5. Aside from
these, $R_1$ is the full $q$ plane. 

\subsection{$L_y=4$}

For this case it is again convenient to give our results in terms of a
generating function $\Gamma(sq_d(L_y=4),PBC_y,FBC_x,q,x)$.  We find 
$d_{\cal D}=3$, $d_{\cal N}=2$, and (using the abbreviation $sqd4cyl$ for this
strip) 
\beq
b_{sqd4cyl,1}=-q^4+16q^3-104q^2+316q-372
\label{bsqd4cyl_1}
\eeq
\beq
b_{sqd4cyl,2}=(q-3)(5q^4-74q^3+422q^2-1100q+1109)
\label{bsqd4cyl_2}
\eeq
and
\beq
b_{sqd4cyl,3}=-(q-2)(q-3)^2(2q^2-16q+33) \ . 
\label{bsqd4cyl_3}
\eeq
Defining 
\beq
A_{sqd4cyl,j}=q(q-1)(q-2)(q-3) \bar A_{sqd4cyl,j}
\label{asqd4factorcyl}
\eeq
we calculate
\beq
\bar A_{sqd4cyl,0}=q^4-14q^3+79q^2-210q+220
\label{asqd4cyl_0}
\eeq
\beq
\bar A_{sqd4cyl,1}=-(5q^5-79q^4+501q^3-1586q^2+2485q-1513)
\label{asqd4cyl_1}
\eeq
and
\beq
\bar A_{sqd4cyl,2}=(q-3)(q^2-3q+3)(2q^2-16q+33) \ . 
\label{asqd4cyl_2}
\eeq
Writing 
\beq
{\cal D}_{sqd4cyl} = \prod_{j=1}^3 (1-\lambda_{sqd4cyl,j} x)
\label{dk4ly3cyldenfac}
\eeq
we have 
\beq
W=(\lambda_{sqd4cyl,j,max})^{1/4} \quad {\rm for} \quad q \in R_1
\label{wr14cyl}
\eeq
where $\lambda_{sqd4cyl,j,max}$ is the $\lambda_{sqd4cyl,j}$ in
(\ref{dk4ly3cyldenfac}) with maximal magnitude in this region.

\begin{figure}[[hbtp]
\centering
\leavevmode
\epsfxsize=4.0in
\begin{center}
\leavevmode
\epsffile{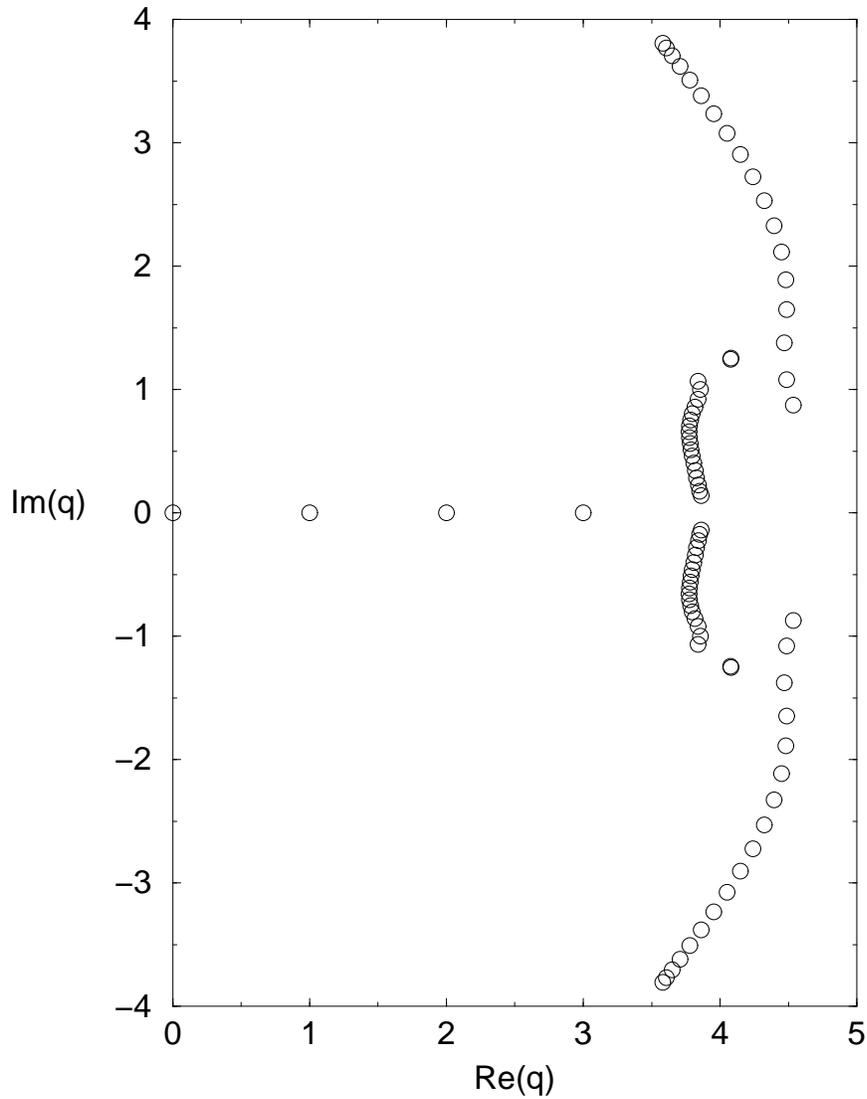}
\end{center}
\caption{\footnotesize{Chromatic zeros for the $4 \times L_x$
cylindrical strip of the $sq_d$ lattice, i.e. with $(PBC_y,FBC_x)$ boundary
conditions with $L_x=20$ (i.e., $n=80$ vertices).}}
\label{k4py4}
\end{figure}

Chromatic zeros for the $4 \times L_x$ strip of the $sq_d$ lattice with
cylindrical boundary conditions are shown in Fig. \ref{k4py4} for $L_x=20$,
i.e. $n=80$.  Again, the length $L_x$ is sufficiently great that these give a
good idea of the location of the asymptotic curve ${\cal B}$.  Since ${\cal B}$
does not cross the real $q$ axis, there is no $q_c$ for this case.

\section{Strip with $L_{\lowercase{y}}=3$ and $(PBC_{\lowercase{y}},(T)PBC_{\lowercase{x}})$}

We consider here the $sq_d$ strip with $L_y=3$ and torus boundary conditions
$(PBC_y,PBC_x)$.  By the same method as mentioned above, e.g., calculating the
associated adjacency matrices and showing that they are the same, it follows
that for a given $L_x$, this strip with torus boundary conditions is identical
to the corresponding $L_y=3$ strip with Klein bottle boundary conditions
$(PBC_y,TPBC_x)$:
\beq
G(sq_d,L_y=3,L_x,PBC_y,PBC_x)=G(sq_d,L_y=3,L_x,PBC_y,TPBC_x)
\label{gtorkleinly3}
\eeq
Since there are 3 vertices on the vertical slice, and hence $3!$ permutations 
that yield identical graphs, it follows, as explained above, that the sum of 
the eigenvalue multiplicities for each individual strip contributes only $1/3!$
of this total:
\beq
C(sq_d,L_y=3,PBC_y,(T)PBC_x) = \frac{1}{3!}P(C_3,q)=\frac{1}{6}q(q-1)(q-2)
\label{csumtorusly3}
\eeq
These strips have the chromatic number 
\beq
\chi(sq_d(L_y=3)_m,PBC_y,(T)PBC_x) = \cases{ 6 & for even $m \ge 4$ \cr 7 & for
odd $m \ge 7$}
\label{chik4torus}
\eeq
For $m=2$ and $m=3$, the $L_y=3$ $sq_d$ torus strip degenerates to 
$K_6$ and $K_9$, respectively, with $\chi(K_p)=p$ as before; for $m=5$ this
strip has $\chi=8$.  The fact that the values of $\chi$ for the
$L_y=3$ torus strip of the $sq_d$ lattice in eq. (\ref{chik4torus}) and the
special cases noted are larger than the value $\chi=4$ for the infinite 2D
$sq_d$ lattice can be ascribed in part to the constraints arising from the
small girth of the triangular transverse cross section of these strips. 

We calculate the chromatic polynomial by iterated use 
of the deletion-contraction theorem, via a generating function approach 
\cite{strip,hs}.  From this we obtain the chromatic polynomial in the form 
(\ref{pgsum}) via the general formulas in \cite{hs} and obtain
\beqs
& & P(sq_d(L_y=3)_m,PBC_y,(T)PBC_x,q) = \frac{1}{6}q(q-1)(q-5)(-6)^m 
+ \frac{1}{2}q(q-3)[6(q-5)]^m \cr\cr
& & + (q-1)\Bigl [-3(q-4)(q-5) \Bigr ]^m + \Bigl [(q-3)(q-4)(q-5)
\Bigr ]^m \ . 
\label{pk4torus}
\eeqs
Thus, $N_\lambda=4$ for this strip. The labelling of the coefficients $c_j$ 
and terms $\lambda_j$ in eq. (\ref{pk4torus}) is consecutive. 
Explicitly calculating the sum of the coefficients in the chromatic 
polynomial (\ref{pk4torus}), 
one sees that the result agrees with (\ref{csumtorusly3}).  It is interesting
that the coefficients that occur in (\ref{pk4torus}) are the same as a subset
of the coefficients that occur in the chromatic polynomial for the $L_y=3$
cyclic strip. 
We find that the degrees in $x$ of the numerator and denominator of the
generating function for this strip graphs are 2 and 4, so that 
all of the $\lambda_j$'s in $d_{\cal D}=4$ contribute to $P$. 

\begin{figure}[p]
\centering
\leavevmode
\epsfxsize=4.0in
\begin{center}
\leavevmode
\epsffile{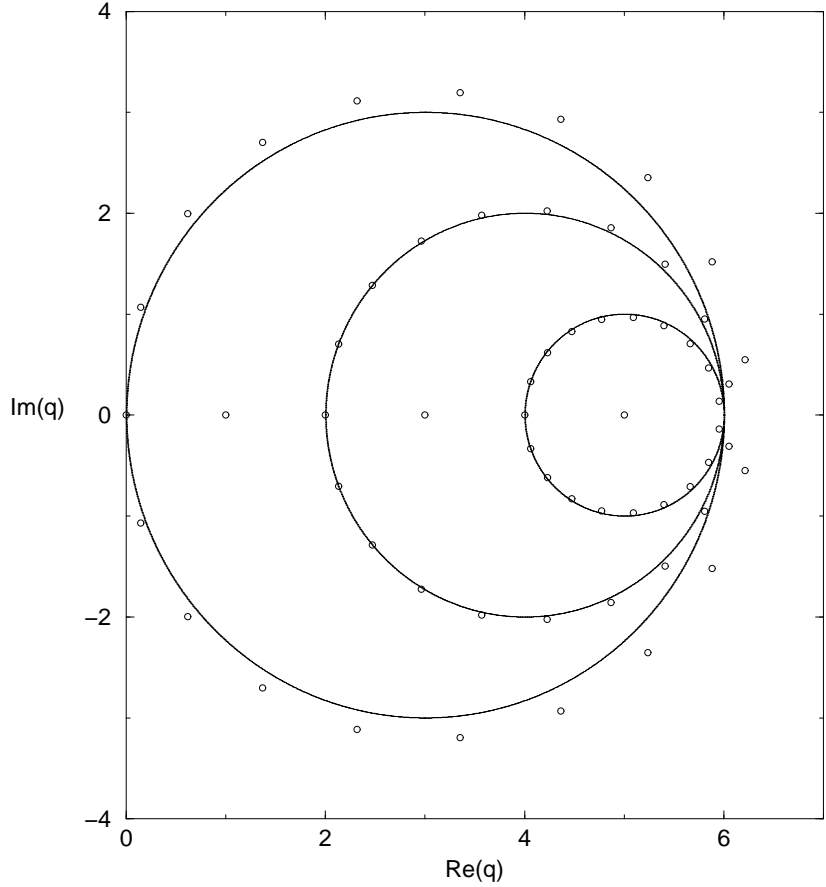}
\end{center}
\caption{\footnotesize{Locus ${\cal B}$ for $W$ for the $3 \times \infty$ 
strip of the $sq_d$ lattice with $(PBC_y,(T)PBC_x)=$ torus or Klein bottle
boundary conditions. Chromatic zeros are shown for
the case $L_x=20$ (i.e., $n=60$).}}
\label{k3k6}
\end{figure}

The nonanalytic locus (boundary) ${\cal B}$ is shown in Fig. \ref{k3k6} and 
consists of three circles that osculate at $q_c$, where
\beq
q_c(sq_d(L_y=3),PBC_y,(T)PBC_x)=6
\label{qcly3}
\eeq
namely, 
\beq
{\cal B}: \{|q-3|=3 \} \ \cup \ \{|q-4|=2 \} \ \cup \ \{|q-5|=1 \} \ . 
\label{bk4ly3torus}
\eeq
Thus, as was true for the $L_y=2$ cyclic strip, $q_c$ is a tacnode. 
Evidently, ${\cal B}$ crosses the real axis at $q=0,2$, and 4 as well as at
$q_c$.  This locus ${\cal B}$ divides the $q$ plane into four regions: (i)
the outermost region $R_1$, including the real intervals $q \ge 6$ and $q \le
0$; (ii) region $R_2$, the interior of the smallest circle, $|q=5|=1$,
containing the real interval $4 \le q \le 6$; (iii) region $R_3$, the interior
of the circle $|q-4|=2$ minus the disk $|q-5|=1$ comprising $R_2$ and including
the real interval $2 \le q \le 4$; and (iv) region $R_4$, the interior of the
largest circle, $|q-3|=3$ minus the disk $|q-4|=2$ and including the real
interval $0 \le q \le 2$.  We have 
\beq 
W=\Bigl [(q-3)(q-4)(q-5)\Bigr ]^{1/3} \quad {\rm for} \quad q \in R_1 \ . 
\label{wk4ly3torusr1}
\eeq
The fact that this coincides with the $W$ calculated for the corresponding
strip with $(PBC_y,FBC_x)$ (see eq. (\ref{wk4ly3pbcyfbcx})) is a general
result \cite{wcy,pm,bcc}.  For the other regions we have 
\beq
|W|=6^{1/3} \quad {\rm for} \quad q \in R_2
\label{w4ly2torusr2}
\eeq
\beq
|W|=|6(q-5)|^{1/3} \quad {\rm for} \quad q \in R_3
\label{w4ly2torusr3}
\eeq
and
\beq
|W|=|3(q-4)(q-5)|^{1/3} \quad {\rm for} \quad q \in R_4 \ . 
\label{w4ly2torusr4}
\eeq
Evidently, for all of these strips, the locus ${\cal B}$ has support for 
$Re(q) \ge 0$.

It is of interest to comment further on the $q_c$ values for (the $L_x \to
\infty$ limits of) these strips of the $sq_d$ lattice.  In our previous exact
calculations of chromatic polynomials for various strip graphs of regular
lattices and the resultant $W$ functions for their $L_x \to \infty$ limits, it
was found that if one uses free transverse boundary conditions and periodic
longitudinal boundary conditions the value of $q_c$ for a given family is a
non-decreasing function of $L_y$.  Our results for the strips of the $sq_d$
lattice exhibit the same behavior.  Hence, our finding that $q_c \simeq 4.25$
for the ($L_x \to \infty$ limit of the) $L_y=3$ cyclic/M\"obius strip graph
suggests that $q_c$ for the infinite 2D $sq_d$ lattice, which we denote
$q_c(sq_d)$, is greater than 4.25.  Note that we cannot use our finding that
$q_c=6$ for the $L_x \to \infty$ limit of the $L_y=3$ torus/Klein bottle graph
of the $sq_d$ lattice to suggest that $q_c(sq_d)$ might be 6 because we have
previously obtained exact solutions for $W$ that show that $q_c$ is not, in
general, a non-decreasing function of $L_y$ on strip graphs with periodic
transverse boundary conditions \cite{strip,t}.  For example, from exact
results, we have found that for the $L_x \to \infty$ limit of strips of the
triangular lattice with cylindrical boundary condition $(PBC_y,FBC_x)$, $q_c=4$
for $L_y=4$ \cite{strip}, while $q_c \simeq 3.28$ for $L_y=4$ and $q_c \simeq
3.25$ for $L_y=5$ \cite{t}.  For the square, triangular, and $sq_d$ lattices,
constructed, say, as the $L_x, \ L_y \to \infty$ limits of open rectangular
sections, one has the chromatic numbers $\chi=2,3$, and 4, respectively.  Now
the Potts antiferromagnet has a zero-temperature critical point at $q=3$ on the
square lattice \cite{lieb}, \cite{schick},\cite{bax82} and at $q=4$ on the
triangular lattice \cite{baxter} respectively (which should be independent of
boundary conditions used in taking the thermodynamic limit), corresponding to
the values $q_c(sq)=3$ and $q_c(tri)=4$.  These results are consistent with the
possibility that $q_c(sq_d)=5$, i.e., the possibility that the $q=5$ Potts
antiferromagnet has a $T=0$ critical point on the $sq_d$ lattice.  However,
there is not a 1-1 correspondence between chromatic number and $q_c$; for
example, the kagom\'e lattice (again constructed, say, as the limit of a finite
section with free transverse and longitudinal boundary conditions) has $\chi=3$
like the triangular lattice, but $q_c=3$, in contrast to the $q_c=4$ value for
the triangular lattice.  If, indeed, $q_c(sq_d)=5$, this would also mean that
$q_c$ for an infinite-length, finite-width strip could be larger than $q_c$ for
the full infinite lattice since we have obtained $q_c=6$ in eq. (\ref{qcly3})
from our exact solution for the chromatic polynomial for the strip with torus
boundary conditions above.

\section{Rigorous Lower Bounds on $W$ and Approach to the Infinite-Width Limit}

Here we present rigorous lower bounds on $W$ for strip graphs and show that
these are very close to the actual values obtained from our exact solutions and
hence serve as very good approximations to the actual $W$ functions.  Using our
exact solutions and these approximations, we then determine how, for a given
$q$, the values of $W$ for the infinite-length, finite-width strips approach
the value for the infinite 2D $sq_d$ lattice as the strip width $L_y$ gets
large. 

As discussed in \cite{pm,bcc}, a general result for a
given type of strip graph $G_s$ is 
\beq
W(G_s(L_y),BC_y,FBC_x,q)=W(G_s(L_y),BC_y,PBC_x,q) \quad {\rm for} \quad q \ge 
q_c(G_s(L_y),BC_y,PBC_x)
\label{wgenrel}
\eeq
where $BC_y=FBC_y$ or $PBC_y$. 
Hence, for example, $W(sq_d(L_y=2),FBC_y,FBC_x,q)=
W(sq_d(L_y=2),FBC_y,PBC_x,q)$ for $q \ge 4$ and 
\beq
W(sq_d(L_y=3),PBC_y,FBC_x,q)=W(sq_d(L_y=3),PBC_y,PBC_x,q) \quad {\rm for} 
\quad q \ge 6
\label{wsqdly3torusrel}
\eeq

We recall that, using coloring matrix methods \cite{b}, Tsai and one of us (RS)
previously derived rigorous upper and lower bounds on $W$ for various 2D
lattices \cite{ww}-\cite{wn}.  It was found that the lower bounds
$W_{\ell. b.}$ were actually very good approximations to the actual $W$ values,
as determined, e.g., by Monte Carlo simulations. This was also seen
analytically from the property that the large-$q$ Taylor series expansions of
$q^{-1}W$ and $q^{-1}W_{\ell. b.}$ coincided to several orders beyond the first
term, which is unity.  The lower bound (\ref{wsqdbound}) was derived as part of
this work.  As noted above, this bound agrees very well with the actual values
of $W$, as determined via Monte Carlo measurements (see Table III of
\cite{w3}).

Using the same methods, we can obtain a lower bound on $W$ for the $sq_d$ 
strips of interest here.  Here we restrict to $q \ge 6$. 
For the case of free transverse boundary conditions and either free or 
periodic longitudinal boundary conditions we obtain the lower bound 
\beq
W(sq_d(L_y),FBC_y,BC_x,q) \ge \frac{[(q-2)(q-3)]^{1-\frac{1}{L_y}}}
{(q-1)^{1-\frac{2}{L_y}}} \ . 
\label{wsqdboundly}
\eeq
For the $sq_d$ strips with periodic transverse boundary conditions and either 
free or periodic longitudinal boundary conditions, we obtain the lower bound 
\beq
W(sq_d(L_y),PBC_y,BC_x,q) \ge A^{\frac{1}{L_y}}
\label{wab}
\eeq
where
\beq
A=\frac{ (q/2)(q-3)2^{L_y}+[(q-2)(q-3)]^{L_y} + (q-1)[2(3-q)]^{L_y}}
{(q-1)^{L_y} + (q-1)(-1)^{L_y}} \ . 
\label{av}
\eeq
Evidently, for large $q$, the lower bound (\ref{wab}) with (\ref{av}) goes over
to (\ref{wsqdbound}) for the infinite lattice.

In Table \ref{tablecomp}, we compare the values of $W$ for $6 \le q \le 10$
from exact solutions for the $L_y=2,3,4$ open strips with the respective 
lower bound (\ref{wsqdbound}).  For each pair of values of $L_y$ and $q$, the 
upper entry is the value of $W$ from the exact solution and the lower entry is
the ratio
\beq
R_{WF}=\frac{W(sq_d(L_y),FBC_y,BC_x,q)}{W(sq_d(L_y),FBC_y,BC_x,q)_{\ell. b.}}
\label{rwo}
\eeq 
where $W(sq_d(L_y),FBC_y,BC_x,q)_{\ell. b.}$ is the lower bound
($\ell. b.$) given by the right-hand side of eq. (\ref{wsqdboundly}) and the
numerator and denominator are independent of the boundary conditions in the
longitudinal direction.  The bound is identical to the exact expression for $W$
for $L_y=2$ and is very close for $L_y=3$ and $L_y=4$.  Thus, just as was found
in our earlier work with Tsai \cite{ww}-\cite{wn}, the lower bound is not just
a bound but also a very accurate approximation to the exact value of $W$,
especially for moderate and large $q$.  Having confirmed this again for the
present type of strip graphs, we use this approximation to study the approach
to the infinite-width limit, i.e. the full infinite 2D $sq_d$ lattice.  In
Table \ref{tableapproach} we show values of $W$ for $6 \le q \le 10$ and
widths $2 \le L_y \le 10$, as compared with the values for $L_y=\infty$,
i.e. for the full 2D $sq_d$ lattice.  For each pair of values $L_y$ and $q$,
the upper entry is the value of $W$ and the lower entry is the ratio of this
value divided by the corresponding value of $W$ for the 2D $sq_d$ lattice.  For
$L_y=2,3,4$, the values of $W$ are from evaluations of the exact solutions,
while for $L_y$ from 5 to 10 they are from the approximation provided by 
(\ref{wab}) with (\ref{av}) and for $L_y=\infty$ they are from the Monte Carlo
measurements given in \cite{w3}.  One sees that for a fixed $L_y$, the
value of $W$ on the infinite-length strip approaches that for the 2D lattice as
$q$ increases.  Clearly, for fixed $q$, the value of $W$ for the
infinite-length strip approaches the value for the 2D lattice as $L_y$
increases, and Table \ref{tableapproach} provides a quantitative picture of
this approach; for example, for $L_y=10$ and a moderate value of $q$ such as 7
or 8, the $W$ values are within several per cent of their 2D lattice values.
We recall that this approach for this type of strip with free transverse
boundary conditions was proved to be monotonic in \cite{w2d}.  

\begin{table}
\caption{\footnotesize{Values of $W$ for infinite-length, finite-width strips
of the $sq_d$ lattice with free transverse boundary conditions, as functions of
$q$, from evaluation of our exact solutions.  For each pair of values of $L_y$
and $q$, the upper entry is the value of $W$ from the exact solution and the
lower entry is the ratio $R_{WF}$.}}
\begin{center}
\begin{tabular}{||c|c|c|c|c|c||}
\hline
$L_y \ \downarrow$ \ \ $q \ \rightarrow$ & 6 & 7 & 8 & 9 & 10  \\ \hline
2  & 3.464 & 4.472 & 5.477 & 6.481 & 7.483 \\
   & 1      & 1      & 1      & 1      & 1      \\ \hline
3  & 3.083 & 4.066 & 5.055 & 6.047 & 7.0415 \\
   & 1.006  & 1.003  & 1.002  & 1.001  & 1.001  \\ \hline
4  & 2.909 & 3.878 & 4.857 & 5.842 & 6.831 \\ 
   & 1.009  & 1.004  & 1.002  & 1.0015 & 1.001  \\ \hline
\end{tabular}
\end{center}
\label{tablecomp}
\end{table}

\begin{table}
\caption{\footnotesize{Values of $W$ for infinite-length, width $L_y$ strips
of the $sq_d$ lattice with $(FBC_y,BC_x)$, as functions of $q$, compared
with the corresponding values for the infinite 2D $sq_d$ lattice.  For each
value of $L_y$ and $q$, the upper entry is the value of $W$ and the lower entry
is the ratio of this value divided by $W(sq_d,q)$ for the 2D $sq_d$ lattice.
For $L_y=2,3,4$, the values of $W$ are taken from the exact solutions; for $5
\le L_y \le 10$ the values are from (\ref{wsqdboundly}).}}
\begin{center}
\begin{tabular}{||c|c|c|c|c|c||}
\hline
$L_y \ \downarrow$ \ \ $q \ \rightarrow$ & 6 & 7 & 8 & 9 & 10  \\ \hline
2  & 3.46  & 4.47  & 5.48  & 6.48  & 7.48  \\
   & 1.42  & 1.33  & 1.27  & 1.23  & 1.20  \\ \hline
3  & 3.08  & 4.07  & 5.055 & 6.05  & 7.04  \\
   & 1.26  & 1.21  & 1.17  & 1.15  & 1.13  \\ \hline
4  & 2.91  & 3.88  & 4.86  & 5.84  & 6.83  \\
   & 1.19  & 1.15  & 1.13  & 1.11  & 1.10  \\ \hline
5  & 2.78  & 3.75  & 4.73  & 5.71  & 6.70  \\
   & 1.14  & 1.11  & 1.10  & 1.08  & 1.07  \\ \hline
6  & 2.71  & 3.68  & 4.65  & 5.63  & 6.62  \\
   & 1.11  & 1.09  & 1.08  & 1.07  & 1.06  \\ \hline
7  & 2.665 & 3.625 & 4.60  & 5.58  & 6.56  \\
   & 1.09  & 1.08  & 1.07  & 1.06  & 1.05  \\ \hline
8  & 2.63  & 3.59  & 4.56  & 5.53  & 6.52  \\
   & 1.075 & 1.07  & 1.06  & 1.05  & 1.04  \\ \hline
9  & 2.60  & 3.56  & 4.53  & 5.50  & 6.48  \\
   & 1.06  & 1.06  & 1.05  & 1.04  & 1.04  \\ \hline
10 & 2.58  & 3.535 & 4.50  & 5.48  & 6.46  \\
   & 1.06  & 1.05  & 1.04  & 1.04  & 1.04  \\ \hline
$\infty$
   & 2.45  & 3.37  & 4.31  & 5.27  & 6.24  \\
   & 1     & 1     & 1     & 1     & 1     \\ \hline 
\end{tabular}
\end{center}
\label{tableapproach}
\end{table}

We next study 
the approach of the values of $W$ on strips of the $sq_d$ lattice with periodic
transverse boundary conditions to their values for the infinite 2D $sq_d$
lattice.  
In Table \ref{tablecompp}, we compare the values of $W$ for $6 \le q \le 10$
from our exact solutions for the $L_y=3,4$ cylindrical strips with the 
respective lower bound (\ref{wab}) with (\ref{av}).  
For each pair of values of $L_y$ and $q$, the 
upper entry is the value of $W$ from the exact solution and the lower entry is
the ratio
\beq
R_{WP}=\frac{W(sq_d(L_y),PBC_y,BC_x,q)}{W(sq_d(L_y),PBC_y,BC_x,q)_{\ell. b.}}
\label{rwp}
\eeq 
where $W(sq_d(L_y),PBC_y,BC_x,q)_{\ell. b.}$ is the lower bound
($\ell. b.$) given by the right-hand side of eq. (\ref{wab}) with (\ref{av}) 
and the
numerator and denominator are independent of the boundary conditions in the
longitudinal direction.  The bound is identical to the exact expression for $W$
for $L_y=3$ and is very close for $L_y=4$.  Hence, as was the case for the open
strip of the $sq_d$ lattice, the lower bound is not just
a bound but also a very accurate approximation to the exact value of $W$,
especially for moderate and large $q$.  Having established this, 
we use this approximation to study the approach
to the infinite-width limit, i.e. the full infinite 2D $sq_d$ lattice.  In
Table \ref{tableapproachp} we show values of $W$ for $6 \le q \le 10$ and
widths $3 \le L_y \le 10$, as compared with the values for $L_y=\infty$,
i.e. for the full 2D $sq_d$ lattice.  For each pair of values $L_y$ and $q$,
the upper entry is the value of $W$ for the infinite cylindrical or torus strip
and the lower entry is the ratio of this
value divided by the corresponding value of $W$ for the 2D $sq_d$ lattice.  For
$L_y=3,4$, the values of $W$ are from evaluations of the exact solutions,
while for $L_y$ from 5 to 10 they are from the approximation provided by bound
(\ref{wsqdboundly}) and for $L_y=\infty$ they are from the Monte Carlo
measurements given in \cite{w3}.  One sees that for a fixed $L_y$, the
value of $W$ on the infinite-length strip approaches that for the 2D lattice as
$q$ increases.  Also, for fixed $q$, the value of $W$ for the
infinite-length strip approaches the value for the 2D lattice as $L_y$
increases.  In \cite{w2d} it was shown that this approach is non-monotonic for
strips of the square and triangular lattice with periodic transverse boundary
conditions and the same is true here.  The approach of the values of $W$ for
the infinite-length finite-width strips to their respective values for the
infinite $sq_d$ lattice is more rapid for the case of periodic transverse
boundary conditions than free transverse boundary conditions.  This is similar
to what was found in \cite{w2d} and is due to the fact that periodic transverse
boundary conditions minimize finite-size artifacts.  Quantitatively, for a
moderate value of the width, such as $L_y=6$, and $q \sim 8$, $W$ is already
within a few parts per thousand of its infinite 2D lattice value. 

\begin{table}
\caption{\footnotesize{Values of $W$ for infinite-length, finite-width strips
of the $sq_d$ lattice with periodic transverse boundary conditions, as
functions of $q$, from evaluation of our exact solutions.  For each pair
of values of $L_y$ and $q$, the upper entry is the value of $W$ from the
exact solution and the lower entry is the ratio $R_{WP}$.}}
\begin{center}
\begin{tabular}{||c|c|c|c|c|c||}
\hline
$L_y \ \downarrow$ \ \ $q \ \rightarrow$ & 6 & 7 & 8 & 9 & 10  \\ \hline
3  & 1.817 & 2.8845 & 3.915 & 4.932 & 5.944 \\
   & 1      & 1      & 1      & 1      & 1      \\ \hline
4  & 2.629 & 3.5005 & 4.412 & 5.348 & 6.300 \\ 
   & 1.024 & 1.014 & 1.009 & 1.006 & 1.004 \\ \hline
\end{tabular}
\end{center}
\label{tablecompp}
\end{table}

\begin{table}
\caption{\footnotesize{Values of $W$ for infinite-length, width $L_y$ strips
of the $sq_d$ lattice with $(PBC_y,BC_x)$, as functions of $q$, compared
with the corresponding values for the infinite 2D $sq_d$ lattice.  For each
value of $L_y$ and $q$, the upper entry is the value of $W$ and the lower entry
is the ratio of this value divided by $W(sq_d,q)$ for the 2D $sq_d$ lattice.
For $L_y=3,4$, the values of $W$ are taken from the exact solutions; for $5
\le L_y \le 10$ the values are from (\ref{wab}) with (\ref{av}).}}
\begin{center}
\begin{tabular}{||c|c|c|c|c|c||}
\hline
$L_y \ \downarrow$ \ \ $q \ \rightarrow$ & 6 & 7 & 8 & 9 & 10  \\ \hline
3  & 1.82  & 2.88  & 3.91  & 4.93  & 5.94  \\
   &0.743  &0.857  &0.908  &0.936  &0.953  \\ \hline
4  & 2.63  & 3.50  & 4.41  & 5.35  & 6.30  \\
   & 1.07  & 1.04  & 1.02  & 1.015 & 1.01  \\ \hline
5  & 2.32  & 3.29  & 4.26  & 5.23  & 6.21  \\
   &0.949  &0.978  &0.989  &0.994  &0.996  \\ \hline
6  & 2.43  & 3.35  & 4.29  & 5.25  & 6.22  \\
   &0.994  &0.994  &0.996  &0.997  &0.998  \\ \hline
7  & 2.39  & 3.33  & 4.28  & 5.25  & 6.22  \\
   &0.976  &0.989  &0.994  &0.996  &0.998  \\ \hline
8  & 2.41  & 3.33  & 4.29  & 5.25  & 6.22  \\
   &0.984  &0.991  &0.995  &0.997  &0.998  \\ \hline
9  & 2.40  & 3.33  & 4.29  & 5.25  & 6.22  \\
   &0.980  &0.990  &0.994  &0.997  &0.998  \\ \hline
10 & 2.40  & 3.33  & 4.29  & 5.25  & 6.22  \\
   &0.982  &0.990  &0.995  &0.997  &0.998  \\ \hline
$\infty$
   & 2.45  & 3.37  & 4.31  & 5.27  & 6.24  \\
   & 1     & 1     & 1     & 1     & 1     \\ \hline 
\end{tabular}
\end{center}
\label{tableapproachp}
\end{table}

One can also compare the results for $W$ with those on the square and 
triangular lattices; this comparison shows the effect of the addition of a
single diagonal bond to each square to go from the square to the triangular
lattice, and then the addition of a second opposite diagonal bond to each
square to go from the triangular to $sq_d$ lattice.  This was done before in 
\cite{w,ww,w3}.

\section{Zero-Temperature Critical Points}

\begin{table}
\caption{\footnotesize{Properties of $P$, $W$, and ${\cal B}$ for strip graphs
$G_s$ of the $sq_d$ lattice.  The properties apply
for a given strip of type $G_s$ of size $L_y \times L_x$; some apply for
arbitrary $L_x$, such as $N_\lambda$, while others apply for the
infinite-length limit, such as the properties of the locus ${\cal B}$.
For the boundary conditions in the $y$ and $x$ directions ($BC_y$,
$BC_x$), F, P, and T denote free, periodic, and orientation-reversed (twisted)
periodic, and the notation (T)P means that the results apply for either
periodic or orientation-reversed periodic. The column denoted eqs. describes
the numbers and degrees of the algebraic equations giving the
$\lambda_{G_s,j}$; for example, $\{6(1),2(2),2(3)\}$ means that there are 
six linear equations, two quadratic equations and two cubic equations.  
The column
denoted BCR lists the points at which ${\cal B}$ crosses the real $q$ axis; the
largest of these is $q_c$ for the given family $G_s$. The notation ``none'' in
this column indicates that ${\cal  B}$ does not cross the real $q$ axis.
Column labelled ``SN'' refers to whether ${\cal B}$ has
\underline{s}upport for \underline{n}egative $Re(q)$, indicated as yes (y) or
no (n).}}
\begin{center}
\begin{tabular}{|c|c|c|c|c|c|c|}
\hline\hline $L_y$ & $BC_y$ & $BC_x$ & $N_\lambda$ & eqs. & BCR & SN
\\ \hline\hline
1 & F    & F & 1   & \{1(1)\}      & none & n       \\ \hline
2 & F    & F & 1   & \{1(1)\}      & none & n       \\ \hline
3 & F    & F & 2   & \{1(2)\}      & none & n       \\ \hline
4 & F    & F & 4   & \{1(4)\}      & none & n       \\ \hline \hline
1 & F    & (T)P & 4   & \{2(1),1(2)\}      & 0, \ 2, \ 3 & n \\ \hline
2 & F    & (T)P & 3 & \{3(1)\}     & 0, \ 2, \ 4 & n \\ \hline
3 & F    & (T)P & 16 & \{6(1),2(2),2(3)\} & 0, \ 2, \ 4, \ 4.25 & n \\ 
\hline\hline
3 & P    & F & 1   & \{1(1)\}      & none & n            \\ \hline
4 & P    & F & 3   & \{1(3)\}      & none & n            \\ \hline\hline
3 & P    & (T)P & 4& \{4(1)\}      & 0, \ 2, \ 4, \ 6 & n \\ \hline\hline
\end{tabular}
\end{center}
\label{sqdprop}
\end{table}

We summarize some of the results obtained in this work in Table \ref{sqdprop}.
Note that the $L_y=1$ strips of the $sq_d$ lattice are equivalent to $L_y=2$
strips of the triangular lattice, as discussed above.  At the value of $q$
where the nonanalytic locus ${\cal B}$ crosses the positive real axis, the
$q$-state Potts antiferromagnet has a zero-temperature critical point
\cite{bcc,a,t}.  From the exact solutions for the chromatic polynomials and $W$
functions, we thus conclude that the $q=2$ Potts (Ising) antiferromagnet has a
$T=0$ critical point on the $L_x \to \infty$ limits of the cyclic/M\"obius
strips with $L_y=2$ and $L_y=3$ as well as the $L_y=3$ strip with torus
(equivalently Klein bottle) boundary conditions.  Note that this involves
frustration owing to the non-bipartite nature of these graphs.  As is generic
for a $T=0$ critical point, the critical singularities are essential, rather
than algebraic, singularities, as we have verified from an explicit transfer
matrix calculation.  The use of strip graphs with periodic or twisted periodic
longitudinal boundary conditions is useful since the crossings of ${\cal B}$ on
the real $q$ axis signal the presence of $T=0$ critical points for the Potts
antiferromagnet at these values of $q$.  Just as was the case with the square
and triangular strips, for which the Ising antiferromagnet also has a $T=0$
critical point \cite{wcy,bcc,a,t}, if one uses free longitudinal boundary
conditions, ${\cal B}$ does not, in general, cross the positive real axis at
$q=2$.  This difference is associated with the noncommutativity in the
definition of $W$, as discussed before \cite{w,bcc,a,t}.  We also find, for
both the cyclic/M\"obius strips and the torus/Klein bottle strips for which we
have calculated exact solutions for $W$, in the nondegenerate cases, $L_y \ge
2$, that the $q$-state Potts antiferromagnet has a $T=0$ critical point at
$q=4$.  Since the $L_x \to \infty$ limit of the cyclic/M\"obius strips can be
carried out with increasing even values of $L_x$, for which the chromatic
number is 4, this zero-temperature critical point is unfrustrated for these
strips.  In contrast, for the $L_x \to \infty$ limit of the torus/Klein bottle
strip, since $min(\chi)=6$, it is frustrated.  Finally, there is formally a
similar critical point at $q=0$.

\section{Conclusions}

In this paper we have presented exact calculations of the zero-temperature
partition function (chromatic polynomial) and $W(q)$, the exponent of the
ground-state entropy, for the $q$-state Potts antiferromagnet with
next-nearest-neighbor spin-spin couplings on strips of the square lattice
strips (equivalently, the nearest-neighbor Potts model on strips of the $sq_d$
lattice) with width $L_y=3$ $L_y=4$ vertices and arbitrarily great length $L_x$
vertices.  Both free and periodic boundary conditions are considered.  In the
$L_x \to \infty$ limit, the resultant $W$ function was calculated.  By
comparing the values of the exact $W$ functions thus obtained for strips with
various widths and boundary conditions versus numerical measurements of $W$ for
the full 2D $sq_d$ lattice, we evaluated the effects of the next-neighbor
spin-spin couplings.  We showed that the $q=2$ (Ising) and $q=4$ Potts
antiferromagnets have zero-temperature critical points on the $L_x \to \infty$
limits of the strips that we studied. With the generalization of $q$ from
${\mathbb Z}_+$ to ${\mathbb C}$, we also determined the analytic structure of
$W(q)$ in the $q$ plane.

\vspace{10mm}

Acknowledgment: The research of R. S. was supported in part at Stony Brook by
the U. S. NSF grant PHY-97-22101 and at Brookhaven by the U.S. DOE contract
DE-AC02-98CH10886.\footnote{\footnotesize{Accordingly, the U.S. government
retains a non-exclusive royalty-free license to publish or reproduce the
published form of this contribution or to allow others to do so for
U.S. government purposes.}}

\vspace{5mm}

Note added:  Since the original submission of this manuscript in Oct., 1999,
several additional related papers have appeared \cite{ta}-\cite{cf}. 

\vspace{4mm}

\section{Appendix} 

\subsection{A Reduction Theorem}

Consider a strip of a given width and length, made up of squares
such that the four vertices of each square are connected to form a complete
graph $K_r$ (so that there are $r-4$ vertices outside of the strip) with some
set of boundary conditions (BC).  We shall denote this strip temporarily as
$G_{sq,K_r,BC}$.  We first show that the chromatic polynomial for
this graph can easily be expressed in terms of the chromatic polynomial for the
corresponding graph with the vertices of each square forming a $K_4$, i.e. such
that the two diagonally opposite vertices of each square are connected by an
edge.  For this purpose we use the intersection theorem from graph theory; this
states that a graph $G$ can be expressed as the union of two subgraphs $G=G_1
\cup G_2$ such that the intersection of these subgraphs is a complete graph,
$G_1 \cap G_2 = K_j$ for some $j$, then $P(G,q)=P(G_1,q)P(G_2,q)/P(K_j,q)$.
Applying this theorem to each square of the above strip, we have 
\beq
P(G_{sq,K_r,BC},q)=\Biggl ( \frac{q^{(r)}}{q^{(4)}} \Biggr )^{N_4}
P(G_{sq,K_4,BC},q) = \Bigl [ \prod_{j=4}^{r-1}(q-j) \Bigr ]^{N_4}
 P(G_{sq,K_4,BC},q)
\label{krk4rel}
\eeq
where $N_4$ denotes the number of squares on the strip and we use the 
standard notation from combinatorics for the ``falling factorial'',
\beq
q^{(s)} \equiv \prod_{j=0}^{s-1}(q-j) \ . 
\label{fallfac}
\eeq
Our results in the text
thus also apply to lattices comprised of squares such that the four vertices 
of each square form a $K_r$ with $r > 4$.

\subsection{$(K_{\lowercase{r}},K_{\lowercase{s}})$ Strips} 

We discuss here a strip of $t$ $K_r$ subgraphs, connected such that 
successive $K_r$ subgraphs intersect on a $K_s$ subgraph, with $1 \le s \le
r-1$, with some longitudinal boundary conditions imposed.  
A member of this general family may be labelled as $(K_r,K_s,t,BC_x)$. Two 
of the simplest longitudinal boundary conditions to
impose are free and cyclic, which we shall denote as $FBC_x$ and $PBC_x$.  The
general $(K_r,K_s,t=m,FBC_x)$ graph has $n=(r-s)m+s$ vertices.

One simple set of families is the cyclic strip $(K_r,K_1,t=m,PBC_x)$.  
These may think of these
graphs as being formed by starting with a circuit graph, $C_m$ and gluing to 
each edge one edge of a complete graph $K_r$.  For these we find
\beq
P((K_r,K_1,m,PBC_x),q) = \Biggl ( \frac{q^{(r)}}{q^{(2)}} \Biggr ) P(C_m,q)
= \Bigl [ \prod_{j=2}^{r-1} (q-j) \Bigr ]^m P(C_m,q)
\label{gluecirc}
\eeq
where $P(C_m,q)$ is the chromatic polynomial for the circuit graph with $m$ 
vertices, given in the introduction. 

Another interesting family is the open strip $(K_r,K_s,m,FBC_x)$.  We calculate
\beq
P((K_r,K_s,m,FBC_x),q)=q^{(s)} \Biggl ( \frac{q^{(r)}}{q^{(s)}} \Biggr )^m = 
q^{(s)} \Bigl [ \prod_{j=s}^{r-1}(q-j) \Bigr ]^m \ . 
\label{prsfbc}
\eeq
Hence, in the $m \to \infty$ limit,
\beq
W=\Biggl ( \frac{q^{(r)}}{q^{(s)}} \Biggr )^{\frac{1}{r-s}} =
 \Bigl [ \prod_{j=s}^{r-1}(q-j) \Bigr ]^{\frac{1}{r-s}}
\label{wkrksfbc}
\eeq
and ${\cal B}=\emptyset$.  

Other cases are more complicated.  For a strip of $K_3$'s, i.e., triangles, in
addition to free and periodic (cyclic) boundary conditions, one also has the
possibility of twisted periodic boundary conditions, which form a M\"obius
strip.  The chromatic polynomial for the $(K_3,K_2,t,PBC_x)$ strip for even $t$
was given in \cite{wcy} (see also \cite{matmeth}); in this case, the strip can
be regarded as being formed from a strip of $m$ squares with a bond added to
each square connecting the lower left to upper right vertex of the square, so
that $t=2m$.  The chromatic polynomial for the corresponding M\"obius strip,
$(K_3,K_2,t,TPBC_x)$ with even $t$ was also given in \cite{wcy}.  For the $t
\to \infty$ limit, the $W$ function and associated nonanalytic locus ${\cal B}$
were given in \cite{wcy}.  The chromatic polynomials for the corresponding
strip with odd $t$, viz., $(K_3,K_2,t,PBC_x)$ and $(K_3,K_2,t,TPBC_x)$, were
given in \cite{t}; the $t \to \infty$ function $W$ and locus ${\cal B}$ are the
same for both the cyclic and M\"obius strips as $t$ increases through even or
odd values.

\begin{figure}[h]
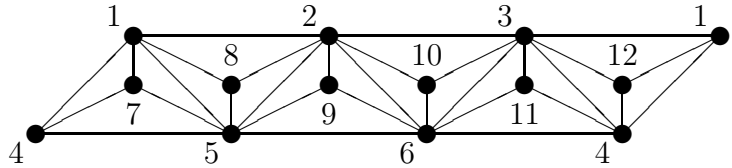

\caption{\footnotesize{Illustrative strip graph comprised of $K_4$ subgraphs
intersecting on common edges ($K_2$'s), of a type different than that shown in
Fig. 1(a).}}
\label{K4}
\end{figure}

In the text we have considered strips in which each square forms a $K_4$
subgraph.  There are actually different classes of strips with $t$ repeated 
$K_4$ subgraphs, in the notation introduced above.  An illustration of these 
is given in Fig. \ref{K4}. 

\vspace{12mm}

\hspace*{3cm}
\begin{picture}(70,10)
\multiput(0,0)(20,0){4}{\circle*{2}}
\multiput(10,5)(10,0){6}{\circle*{2}}
\multiput(10,10)(20,0){4}{\circle*{2}}
\put(0,0){\line(1,0){60}}
\put(10,10){\line(1,0){60}}
\multiput(0,0)(20,0){4}{\line(1,1){10}}
\multiput(10,10)(20,0){3}{\line(1,-1){10}}
\multiput(0,0)(20,0){3}{\line(2,1){10}}
\multiput(20,0)(20,0){3}{\line(-2,1){10}}
\multiput(10,5)(20,0){3}{\line(0,1){5}}
\multiput(10,10)(20,0){3}{\line(2,-1){10}}
\multiput(30,10)(20,0){3}{\line(-2,-1){10}}
\multiput(20,0)(20,0){3}{\line(0,1){5}}
\put(-2,-2){\makebox(0,0){4}}
\put(18,-2){\makebox(0,0){5}}
\put(38,-2){\makebox(0,0){6}}
\put(58,-2){\makebox(0,0){4}}
\put(10,2){\makebox(0,0){7}}
\put(20,8){\makebox(0,0){8}}
\put(30,2){\makebox(0,0){9}}
\put(40,8){\makebox(0,0){10}}
\put(50,2){\makebox(0,0){11}}
\put(60,8){\makebox(0,0){12}}
\put(8,12){\makebox(0,0){1}}
\put(28,12){\makebox(0,0){2}}
\put(48,12){\makebox(0,0){3}}
\put(68,12){\makebox(0,0){1}}
\end{picture}

\vspace{10mm}

As an example, we calculate 
\beq
P((K_4,K_2,t,BC_x),q) = (q-3)^t \ P((K_3,K_2,t,BC_x),q)
\label{k42}
\eeq
where $BC_x=PBC_x$ or $TPBC_x$, respectively and the $(K_3,K_2,t,PBC_x)$ and 
$(K_3,K_2,t,TPBC_x)$ strips are the cyclic and M\"obius triangular-lattice 
strips mentioned above.  Thus, explicitly, for even $t=2m$, 
\beq
P((K_4,K_2,t=2m,PBC_x),q)=(q-3)^{2m}\biggl [ (q^2-3q+1) + [(q-2)^2]^m + 
(q-1)\Bigl [(\lambda_{t2,3})^m + (\lambda_{t2,4})^m \Bigr ] \biggr ] 
\label{p42cycnteven}
\eeq
\beq
P((K_4,K_2,t=2m,TPBC_x),q)=(q-3)^{2m}\biggl [-1 +[(q-2)^2]^m -
(q-1)(q-3)\frac{\Bigl [ (\lambda_{t2,3})^m - (\lambda_{t2,4})^m \Bigr ]}{
\lambda_{t2,3}-\lambda_{t2,4}} \biggr ]
 \label{p42mbnteven}
\eeq
where $\lambda_{t2,j}$, $j=3,4$, were defined in eqs. (\ref{lamtly234}) above. 

For the case of odd $t=2m+1$, we have 
\beqs
& & P((K_4,K_2,t=2m+1,PBC_x),q)=(q-3)^{2m+1}\Biggl [ -(q^2-3q+1) + 
(q-2)[(q-2)^2]^m + \cr\cr
& & \frac{1}{2}(q-1)(q-3)\Biggl [\Bigl ( (\lambda_{t2,3})^m + 
(\lambda_{t2,4})^m \Bigr ) + 
\frac{\Bigl ( (\lambda_{t2,3})^m - (\lambda_{t2,4})^m \Bigr )}{
\lambda_{t2,3}-\lambda_{t2,4}} \Biggr ] \ \Biggr ] 
\label{p42cycntodd}
\eeqs
\beqs
& & P((K_4,K_2,t=2m+1,TPBC_x),q)= (q-3)^{2m+1}\Biggl [ 1 + (q-2)[(q-2)^2]^m+
\cr\cr
& & \frac{1}{2}(1-q)\Biggl [ \Bigl ( (\lambda_{t2,3})^m+ (\lambda_{t2,4})^m
\Bigr ) + (9-4q)\frac{\Bigl ( (\lambda_{t2,3})^m - (\lambda_{t2,4})^m \Bigr )}
{\lambda_{t2,3}-\lambda_{t2,4}} \Biggr ] \ \Biggr ] \ .
\label{p42mbntodd}
\eeqs

\vfill
\eject
\end{document}